\documentclass[reprint,aps,prl,nofootinbib]{revtex4-1}
\usepackage{graphicx}
\usepackage[utf8]{inputenc}
\usepackage{amsmath}
\usepackage{natbib}
\usepackage{bm}
\usepackage{siunitx}
\usepackage{placeins}
\usepackage[caption=false]{subfig}
\usepackage{amsmath} 
\graphicspath{{./images/}}

\begin{document}
\title{Quantitative disentanglement of spin Seebeck, proximity-induced and intrinsic anomalous Nernst effect in NM/FM bilayers}
\author{Panagiota Bougiatioti$^1$, Christoph Klewe$^{1,2}$, Daniel Meier$^1$, Orestis Manos$^1$, Olga Kuschel$^3$, Joachim Wollschl\"ager$^3$, Laurence Bouchenoire$^{4,5}$, Simon D. Brown$^{4,5}$, Jan-Michael Schmalhorst$^1$, G\"unter~Reiss$^1$, and Timo~Kuschel$^{1,6}$\email{Electronic mail: pbougiatioti@physik.uni-bielefeld.de}}
\affiliation{$^1$\mbox{Center for Spinelectronic Materials and Devices, Department of Physics,}\\
\mbox{Bielefeld University, Universit\"atsstra\ss e 25, 33615 Bielefeld, Germany}\\
$^2$\mbox{Advanced Light Source, Lawrence Berkeley National Laboratory, Berkeley, CA 94720, USA}\\
$^3$\mbox{Department of Physics and Center of Physics and Chemistry of New Materials,} \\
\mbox{Osnabrück University, Barbarastrasse 7, 49076 Osnabrück, Germany}\\
$^4$\mbox{XMaS, European Synchrotron Radiation Facility, Grenoble, 38043, France}\\
$^5$\mbox{Department of Physics, University of Liverpool, Liverpool L69 7ZE, UK}\\
$^6$\mbox{Physics of Nanodevices, Zernike Institue for Advanced Materials,}\\
\mbox{University of Groningen, Nijenborgh 4, 9747 AG Groningen, The Netherlands}\\}
\date{\today}


\begin{abstract}
 
We identify and investigate thermal spin transport phenomena in sputter-deposited Pt/NiFe$_2$O$_{\textrm{4-x}}$ ($4\geq x \geq 0$) bilayers. We separate the voltage generated by the spin Seebeck effect from the anomalous Nernst effect contributions and even disentangle the intrinsic anomalous Nernst effect (ANE) in the ferromagnet (FM) from the ANE produced by the Pt that is spin polarized due to its proximity to the FM. Further, we probe the dependence of these effects on the electrical conductivity and the band gap energy of the FM film varying from nearly insulating NiFe$_2$O$_{\textrm{4}}$ to metallic Ni$_{33}$Fe$_{67}$. A proximity-induced ANE could only be identified in the metallic Pt/Ni$_{33}$Fe$_{67}$ bilayer in contrast to Pt/NiFe$_2$O$_{\textrm{x}}$ ($x>0$) samples. This is verified by the investigation of static magnetic proximity effects via x-ray resonant magnetic reflectivity.
 
\end{abstract}

\maketitle

In the emerging fields of spintronics \cite{Hoffmann:2015} and spin caloritronics \cite{Bauer:2012} phenomena such as the spin Hall effect (SHE) \cite{Hirsch:1999} and the spin Seebeck effect (SSE) \cite{Uchida:2010,Uchida:2014} enable the generation, manipulation and detection of spin currents in ferro(i)magnetic insulators (FMI). The most common path to detect a spin current is to use a normal metal (NM) with a large spin Hall angle, such as Pt \cite{Liu:2011}, Ta \cite{Liu:2012}, Pd \cite{Tang:2013} and W \cite{Pai:2012} on top of an FM material. The inverse spin Hall effect (ISHE) \cite{Saitoh:2006} then leads to the conversion of the spin current into a transverse charge voltage in the NM.
\vspace{-0.06em}

Pt is employed frequently for generating and detecting pure spin currents, if adjacent to an FMI, although the possibility of magnetic proximity effects (MPEs) has to be taken into account. Due to its close vicinity to the Stoner criterion \cite{Stoner:1938} the FM can potentially generate a Pt spin polarization at the interface. Consequently, this might induce additional parasitic effects preventing the correct interpretation of the measured ISHE voltage. Therefore, a comprehensive investigation regarding the magnetic properties of the NM/FM interface is required to distinguish the contributions of such parasitic voltages from the ISHE voltage generated by a pure spin current.
\vspace{-1em}

In the case of SSE, the driving force for the spin current in the FM or FMI is a temperature gradient. When a spin current is generated parallel to a temperature gradient, it is generally attributed to the longitudinal spin Seebeck effect (LSSE) \cite{Uchida:2010,Uchida:2014}. However, when using the ISHE in an adjacent NM for the spin current detection, not only a proximity-induced ANE \cite{Huang:2012} can contaminate the LSSE signal, but also an additional intrinsic ANE contribution could be present in case of studying ferromagnetic metals (FMMs) or semiconducting ferro(i)magnets \cite{Meier:2013,Ramos:2013}. Mainly NM/FMI bilayers have been investigated, while LSSE studies on NM/FMM are quite rare.\par

However, Ramos \textit{et al}. \cite{Ramos:2013, Ramos:2014,Ramos:2015,Ramos:2016} and Wu \textit{et al}. \cite{Wu:2014} individually investigated the LSSE in magnetite, which is conducting at room temperature (RT) and, thus, has an intrinsic ANE contribution. They identified the LSSE in Pt/Fe$_3$O$_4$ \cite{Ramos:2013} and CoFeB/Fe$_3$O$_4$ bilayers \cite{Wu:2014} by using temperatures below the conductor-insulator transition of magnetite (Verwey transition at 120$\,$K) in order to exclude any intrinsic ANE contribution. Ramos \textit{et al}. further investigated the ANE in bulk magnetite without any Pt \cite{Ramos:2014} and concluded that the ANE contributions for Pt/Fe$_3$O$_4$ bilayers and multilayers should be quite small \cite{Ramos:2015,Ramos:2016}. In addition, Lee \textit{et al}. \cite{Lee:2015} and Uchida \textit{et al}. \cite{Uchida:2015,Uchida:2016} discussed that in Pt/FMM multilayers both LSSE and ANE contribute, but did not disentangle the effects quantitatively. Hence, a clear quantitative disentanglement of the LSSE in the FMM \cite{FOOTNOTE}, the intrinsic ANE in the FMM, and the proximity-induced ANE in the NM is still pending.

Some groups used Cu or Au interlayers to suppress the MPE in NM/FMM bilayers \cite{Kikkawa:2013PRL,Xu:2014,Miao:2016}. However, a promising technique to distinguish between LSSE and proximity-induced ANE was first proposed by Kikkawa \textit{et al}. \cite{Kikkawa:2013PRL,Kikkawa:2013PRB}. In their study, the voltage measured transverse to the thermal gradient in in-plane magnetized (IPM) and out-of-plane magnetized (OPM) configurations, leads to the sufficient separation of the aforestated contributions. So far, this technique was only used to study the proximity-induced ANE in NM/FMI bilayers. It has not yet been applied to fully conducting NM/FMM bilayers for the separation of the LSSE and ANE contributions in the FMM. In our work, we extend this technique to identify all three contributions quantitatively: LSSE, intrinsic ANE in the FM, and proximity-induced ANE. We will use this separation for investigating these effects in Pt on different FM materials such as nearly-insulating NiFe$_2$O$_4$, semiconducting-like NiFe$_2$O$_{\textrm{x}}$ ($4>x>0$), and metallic Ni$_{33}$Fe$_{67}$.

To confirm or exclude any possible static MPE at the interface of a Pt/FM hybrid structure, element-selective x-ray resonant magnetic reflectivity (XRMR) has been used due to its sensitivity to magnetic moments at interfaces \cite{Macke:2014,Kuschel:2015}. XRMR measurements were performed at the XMaS beamline BM28 at ESRF (Grenoble, France) \cite{Brown:2001}, at RT. Details for the XRMR technique, experiment and data processing can be found in the Supplemental Materials II \cite{Bou} (including Ref. \cite{Bouchenoire:2003}).

We fabricated the films on MgAl$_2$O$_4$ (MAO) substrates by reactive sputter deposition \cite{Klewe:2014} starting from pure high-resistive NiFe$_2$O$_4$ (NFO) $(\sim160\,$nm) up to the metallic Ni$_{33}$Fe$_{67}$ (10.4$\,$nm) with intermediate  NiFe$_2$O$_{\textrm{x}_1}$ (60$\,$nm) and NiFe$_2$O$_{\textrm{x}_2}$ ($35\,$nm), with $4>x_1>x_2>0$, see Supplemental Materials I \cite{Bou}. Twin FM layers have been prepared with and without Pt in-situ deposited on top, in a range of (2.7-3.5)$\,$nm, by covering one FM layer with a mask to maintain the same deposition conditions for the FM in both samples.\par

\vspace{-0.097em}


\begin{figure}[!ht]
    \centering
    \includegraphics[height=7cm, width=\linewidth]{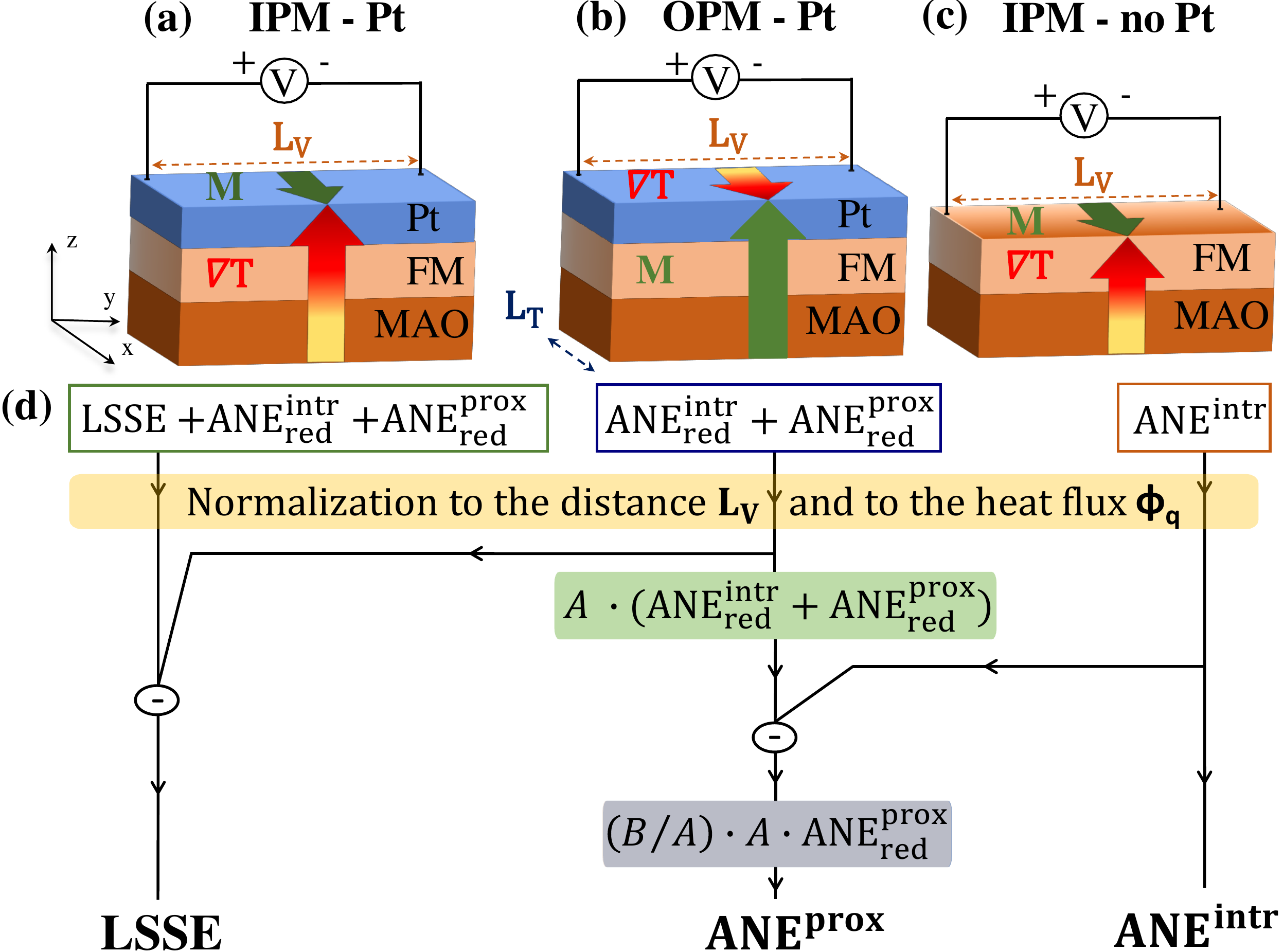}
  \caption{Schematic illustration of (a),(c) in-plane magnetized and (b) out-of-plane magnetized geometries, introducing the temperature gradient $\nabla T$, the magnetization vector $\bm{M}$, the distance between the contacts $L_{\textrm{V}}$ and the total length of the sample $L_{\textrm{T}}$, respectively. (d) Flow chart for the quantitative separation of both ANE contributions from the LSSE voltage. The light green and grey areas correspond to the intermediate steps determining the correction factors $A$ and $B$ respectively, taking into account the reduction of the ANE signal due to the additional Pt layer (spin polarized and/or non-magnetic).}
    \label{fig:geometries}
\end{figure}


 Figures \ref{fig:geometries}(a)-(c) illustrate the measurement geometries that we have employed for the separation of the three effects. In the IPM geometries (Figs. \ref{fig:geometries}(a),(c)) the application of an out-of-plane temperature gradient $\nabla$T in the presence of an in-plane magnetic field along the x-axis induces a transverse voltage along the y-axis. While measuring in this IPM configuration with Pt on top (IPM$\,$-$\,$Pt, Fig. \ref{fig:geometries}(a)) we detect the LSSE voltage together with both ANE contributions, intrinsic and proximity-induced. However, in the IPM geometry without Pt (IPM$\,$-$\,$no$\,$Pt, Fig. \ref{fig:geometries}(c)) we are only sensitive to the intrinsic ANE contribution.\par
 
 The LSSE voltage is determined according to the relation

\vspace{-1em}
\begin{equation}
\bm{E}_{\textrm{ISHE}} = S_{\textrm{SSE}} \bm{J_\textrm{s}} \times \bm{s}
\end{equation}

\noindent where $\bm{E}_{\textrm{ISHE}}$, $S_{\textrm{SSE}}$, $\bm{J_\textrm{s}}$, and $\bm{s}$ denote the electric field induced by ISHE, the SSE coefficient, the spin current which enters the spin detector material and the spin polarization vector, respectively. Moreover, the ANE contribution is described by the relation

\vspace{-1em}
\begin{equation}
\bm{E}_{\textrm{ANE}}  = D_{\textrm{ANE}} \bm\nabla T \times \bm{M}
\end{equation}

\noindent where $\bm{E}_{\textrm{ANE}}$,  $D_{\textrm{ANE}}$, and $\bm{M}$ denote the electric field induced by ANE, the ANE coefficient, and the magnetization vector of the FM, respectively.\par

In the OPM geometry with Pt on top (OPM$\,$-$\,$Pt, Fig. \ref{fig:geometries}(b)), the application of an in-plane temperature gradient $\nabla$T together with an out-of-plane magnetic field generates a transverse voltage attributed to the intrinsic and proximity-induced ANE. In this configuration, the LSSE can not be detected, since no out-of-plane spin current with the proper spin polarization direction is generated \cite{Kikkawa:2013PRL}. One major issue is to consider the reduction of the ANE signal upon a placement of a Pt layer \cite{Ramos:2013}. All ANE contributions measured with Pt on top have in general reduced contributions and this is indicated by the subscript ``red" in Fig. \ref{fig:geometries} and throughout the whole manuscript.\par

Figure \ref{fig:geometries}(d) explains the flow chart for the quantitative disentanglement of the three effects. As a first step, the electric field is calculated from the measured voltages by normalizing to the distance of the electric contacts $L_\textrm{V}$. Then, this electric field is divided by the heat flux $\phi_\textrm{q}$ that runs through the sample. The normalization to the heat flux as suggested by Sola \textit{et al} \cite{Sola:2015, Sola:2016}, allows eliminating the systematic errors due to the thermal surface resistances and thermal contacts resulting in the effective comparison between IPM and OPM configurations as well as in the comparability of our results. Further details  on the heat flux normalization can be found in the Supplemental Materials III \cite{Bou} (including Refs. \cite{Schulz:2002,Nelson:2014}). To estimate the ANE reduction due to the additional Pt layer we used the ratio of conductances $G$ of the NiFe$_2$O$_{\textrm{x}}$ and the Pt in a parallel arrangement \cite{Ramos:2013}

\vspace{-1em}
\begin{equation}
r=\frac{G_{\textrm{NiFe}_{\textrm{2}} \textrm{O}_{\textrm{x}}}}{G_{\textrm{Pt}}}=\frac{\rho_{\textrm{Pt}}}{\rho_{\textrm{NiFe}_{\textrm{2}} \textrm{O}_{\textrm{x}}}}\frac{t_{\textrm{NiFe}_{\textrm{2}} \textrm{O}_{\textrm{x}}}}{t_\textrm{Pt}}
\end{equation}

\noindent with $\rho$: RT resistivity and $t$: thickness of the corresponding layer. The reduced intrinsic ANE signal (ANE$_\textrm{red}^\textrm{intr}$) from the OPM$\,$-$\,$Pt configuration is then corrected by the factor $A=\frac{r+1}{r}$ \cite{Ramos:2013} resulting in the pure ANE$^{\textrm{intr}}=A\,\cdot\,$ANE$_\textrm{red}^\textrm{intr}$. This correction step in our calculations is highlighted by the light green area in Fig. \ref{fig:geometries}(d). 
Combined with the information on the ANE$^\textrm{intr}$ from the IPM$\,$-$\,$no$\,$Pt configuration (cf. Fig. \ref{fig:geometries}(c)), i.e., by subtracting the ANE$^\textrm{intr}$ from the corrected term, this method already yields a qualitative criterion for the existence or absence of proximity-induced ANE in the sample.\par 

For a quantitative evaluation, an additional correction has to be applied to the reduced proximity-induced ANE signal (ANE$_\textrm{red}^\textrm{prox}$) due to the additional non-magnetic Pt layer, while the correction $A$ on the term has to be reversed (see light grey area in Fig. \ref{fig:geometries}(d)).
\begin{figure}[!ht]
    \centering
    \includegraphics[height=6.7cm, width=\linewidth]{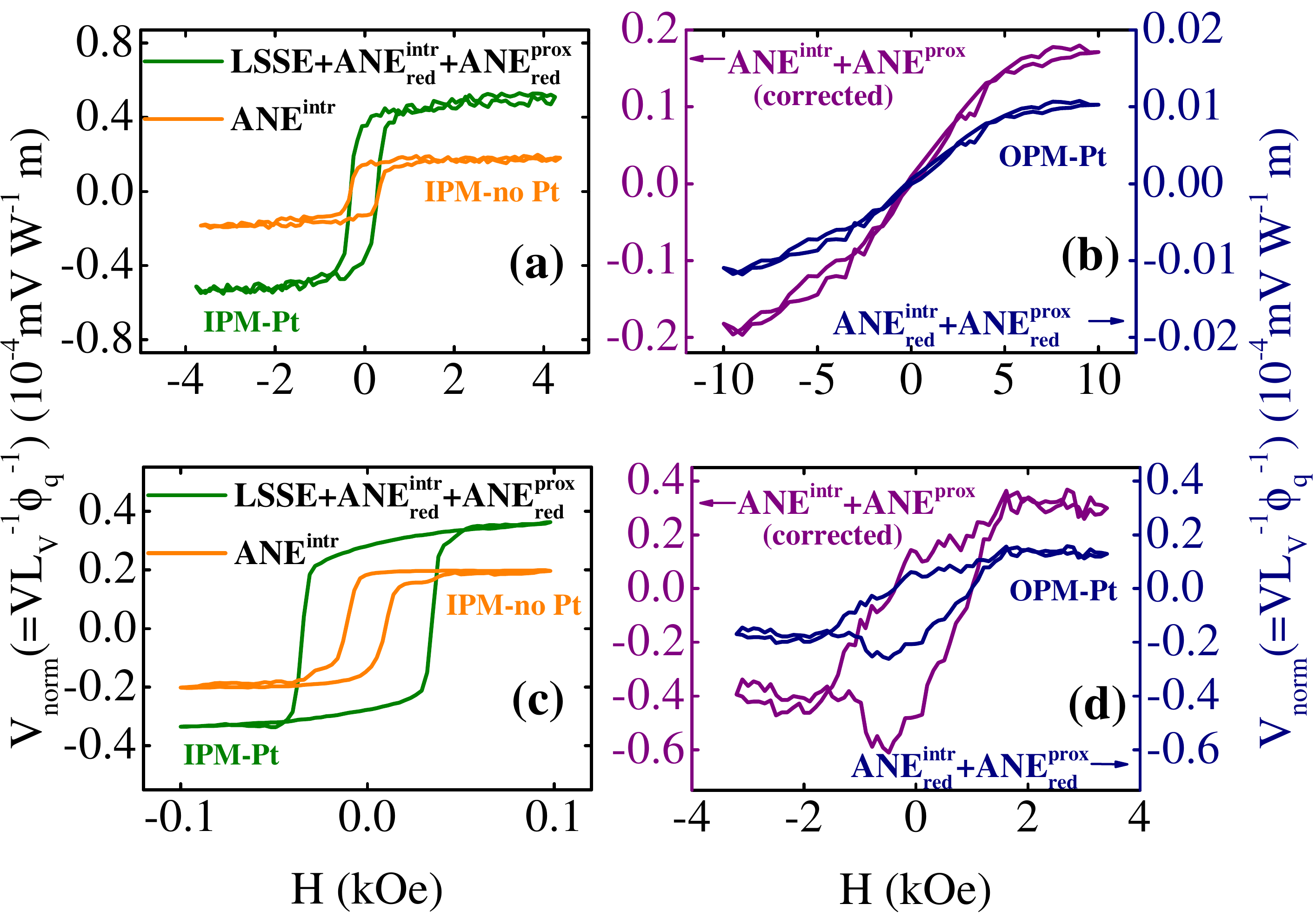}
    \caption{Normalized voltage plotted against the magnetic field strength for (a), (b) Pt/NiFe$_2$O$_{\textrm{x}_{2}}$ and (c), (d) Pt/Ni$_{33}$Fe$_{67}$ bilayers measured in (a), (c) IPM and (b), (d) OPM geometries with the corresponding separation of the ANE contribution (intrinsic and proximity-induced) from the LSSE voltage. ANE$^\textrm{intr}$ + ANE$^\textrm{prox}$ (purple) regards the calculated ANE signal after the implementation of the correction factors $A$ and $B$, which correct the reduction of the measured ANE from the OPM$\,$-$\,$Pt configuration due to the additional Pt layer (spin polarized and/or non-magnetic).}
    \label{fig:loops}
\end{figure}

\FloatBarrier

 The correction factor for the ANE$_\textrm{red}^\textrm{prox}$ is given by $B=\frac{d_{\textrm{I}}+d_{\textrm{II}}}{d_{\textrm{I}}}$ \cite{Ramos:2013}, where $d_{\textrm{I}}$ and $d_{\textrm{II}}$ are the thicknesses of the spin polarized Pt layer and the non-magnetic fraction, respectively, estimated by XRMR. Then, the corrected proximity-induced ANE contribution is denoted as ANE$^\textrm{prox}=(B/A)\,\cdot\,A\,\cdot\,$ANE$_\textrm{red}^\textrm{prox}$. For the polarized and unpolarized fraction of the Pt layers, the same resistivity $\rho_\textrm{Pt}$ was used.

Exemplarily, for the Pt/Ni$_{33}$Fe$_{67}$ (Pt/NiFe$_2$O$_{\textrm{x}_{2}}$) sample the reduction of the ANE$^\textrm{intr}$ is estimated to be 47$\%$ (95$\%$) by using the measured values for the RT resistivity of Pt equal to  $\rho_\textrm{Pt}$ = $1.6\cdot10^{-7}\,\Omega \textrm{m}$ $(1.8\cdot10^{-7}\,\Omega \textrm{m})$ for a Pt film with thickness $t_\textrm{Pt}=3.5\,$nm (3.1$\,$nm) and of the FM equal to $\rho_{\textrm{Ni}_{\textrm{33}}\textrm{Fe}_{\textrm{67}}(\textrm{NiFe}_{\textrm{2}}\textrm{O}_{\textrm{x}_{2}})}$ = $4.2\cdot10^{-7}\,\Omega \textrm{m}$ $(4.5\cdot10^{-5}\,\Omega \textrm{m}) $ for a FM thickness of $t_{\textrm{Ni}_{\textrm{33}}\textrm{Fe}_{\textrm{67}}(\textrm{NiFe}_{\textrm{2}}\textrm{O}_{\textrm{x}_{2}})}$  = 10.4$\,$nm (35$\,$nm). 
Moreover, for the metallic Pt/Ni$_{33}$Fe$_{67}$ bilayer the reduction of the ANE$^\textrm{prox}$ is estimated to be 71$\%$ by considering $d_{\textrm{I}}=1.0\,$nm spin polarized layer of Pt and $d_{\textrm{II}}=2.5\,$nm of non-magnetic Pt. A table with the obtained values for all samples can be found in the Supplemental Materials IV \cite{Bou}. Consequently, the comparison between the voltage signals in the IPM and OPM geometries enables a quantitative separation of the parasitic ANE contributions from the LSSE signal.\par

Figure \ref{fig:loops} illustrates the experimental results for the Pt/NiFe$_2$O$_{\textrm{x}_{2}}$ and Pt/Ni$_{33}$Fe$_{67}$ bilayers. For Pt/NiFe$_2$O$_{\textrm{x}_{2}}$ (Fig. \ref{fig:loops}(a)), the LSSE is the most prominent contribution to the total voltage signal, while for Pt on metallic Ni$_{33}$Fe$_{67}$, the intrinsic ANE and the LSSE are of comparable magnitude (Fig. \ref{fig:loops}(c)). By comparing the difference between the ANE$^\textrm{intr}$ from the IPM$\,$-$\,$no$\,$Pt configuration and the ANE$^\textrm{intr}$ + ANE$^\textrm{prox}$ signals (corrected ANE$_\textrm{red}^\textrm{intr}$ + ANE$_\textrm{red}^\textrm{prox}$ by $A$ and $B$, as explained above) we are able to quantitatively determine the contribution from the proximity-induced ANE. For the non-metallic NiFe$_2$O$_{\textrm{x}_{2}}$ bilayer (Fig. \ref{fig:loops}(a),(b)) no difference can be determined between the saturation values of the ANE$^\textrm{intr}$ data from IPM$\,$-$\,$no$\,$Pt configuration (orange line in Fig. \ref{fig:loops}(a)) and the saturation values of the ANE$^\textrm{intr}$ + ANE$^\textrm{prox}$ signal (corrected OPM$\,$-$\,$Pt data, purple line in Fig. \ref{fig:loops}(b)), which are extracted to be V$^{\textrm{sat}}_{\textrm{norm}}=(0.18\pm0.02)\,10^{-4}\textrm{mV}\textrm{W}^{-1}\textrm{m}$ in both cases. Thus, the ANE$^\textrm{prox}$ is zero and can be neglected for this sample. On the contrary, for the Pt/Ni$_{33}$Fe$_{67}$ bilayer (Fig. \ref{fig:loops}(c),(d)) the ANE$^\textrm{intr}$ + ANE$^\textrm{prox}$ is $(46\pm 3)\%$ larger than the ANE$^\textrm{intr}$ signal unveiling the existence of MPE. Furthermore, for the Pt/NFO bilayer both ANE$^\textrm{intr}$ and ANE$^\textrm{intr}$ + ANE$^\textrm{prox}$ signals are zero confirming the absence of any ANE contribution in the pure Pt/NFO bilayer \cite{Kuschel:2015,Kuschel:2016}.\par


\begin{figure}[!ht]
    \centering
    \includegraphics[height=8cm, width=\linewidth]{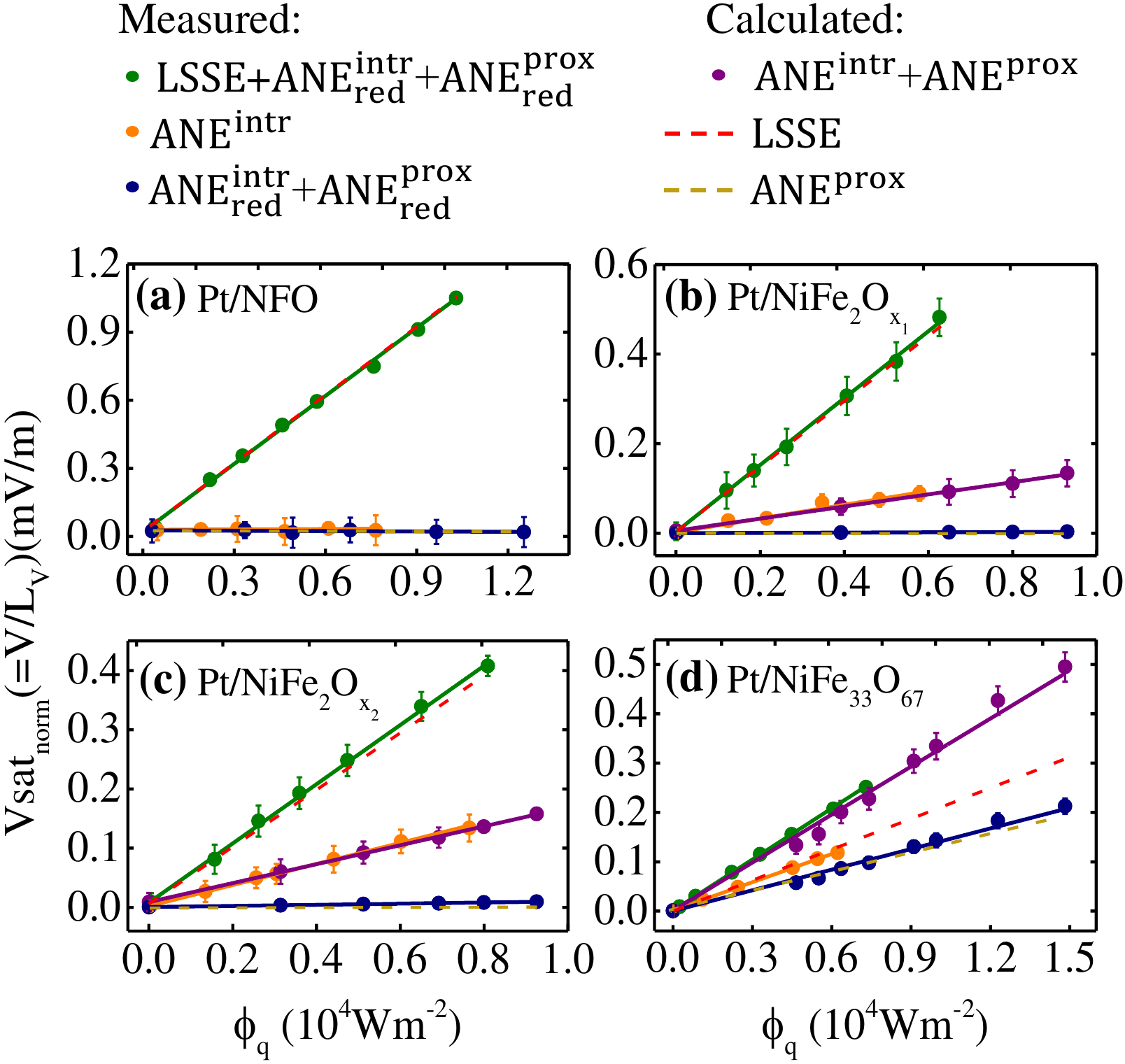}
    \caption{Normalized voltage in saturation against the heat flux for (a) Pt/NFO, (b),(c) Pt/NiFe$_2$O$_{\textrm{x}_{\textrm{1}}/\textrm{x}_{\textrm{2}}}$, and (d) Pt/Ni$_{33}$Fe$_{67}$ samples with the corresponding separation of the ANE contribution (intrinsic and proximity-induced) from the LSSE voltage.}
    \label{fig:slopes}
\end{figure}

Figure \ref{fig:slopes} illustrates the linear dependence of the voltage in saturation on $\phi_\textrm{q}$, normalized to $L_\textrm{V}$ for all samples. The dashed lines are the calculated contributions of the pure LSSE and ANE$^\textrm{prox}$ extracted as described in the diagram of Fig. \ref{fig:geometries}(d) after correcting the reduced ANE signal arising from both the FM and the spin polarized Pt layer. In Fig. \ref{fig:slopes}(a), the zero line contribution of both types of ANE indicates the absence of MPE in Pt/NFO bilayers \cite{Kuschel:2015,Kuschel:2016}. The low amount of mobile charge carriers in the nearly-insulating NFO leads to a vanishing ANE$^\textrm{intr}$ contribution \cite{Meier:2013}.\par

As shown in Figs. \ref{fig:slopes}(a)-(c), the LSSE contribution is dominant for all Pt/NiFe$_2$O$_{\textrm{x}}$ ($x>0$) bilayers that consist of oxides. Furthermore, the absence of any proximity-induced ANE is verified, since no difference between the ANE$^\textrm{intr}$ and the  ANE$^\textrm{intr}$ + ANE$^\textrm{prox}$ can be identified. Additionally, for the Pt/NiFe$_2$O$_{\textrm{x}_{2}}$ bilayer the ANE$^\textrm{intr}$ contribution is 14$\%$ larger than for the Pt/NiFe$_2$O$_{\textrm{x}_{1}}$ bilayer pointing towards its more conducting character. For the Pt/Ni$_{33}$Fe$_{67}$ bilayer (Fig. \ref{fig:slopes}(d)), the enhancement of ANE$^\textrm{intr}$ + ANE$^\textrm{prox}$ due to the metallic character of Ni$_{33}$Fe$_{67}$ and the MPE contribution is clearly displayed. \par

\begin{figure}[!ht]
    \centering
    \includegraphics[height=6.5cm, width=\linewidth]{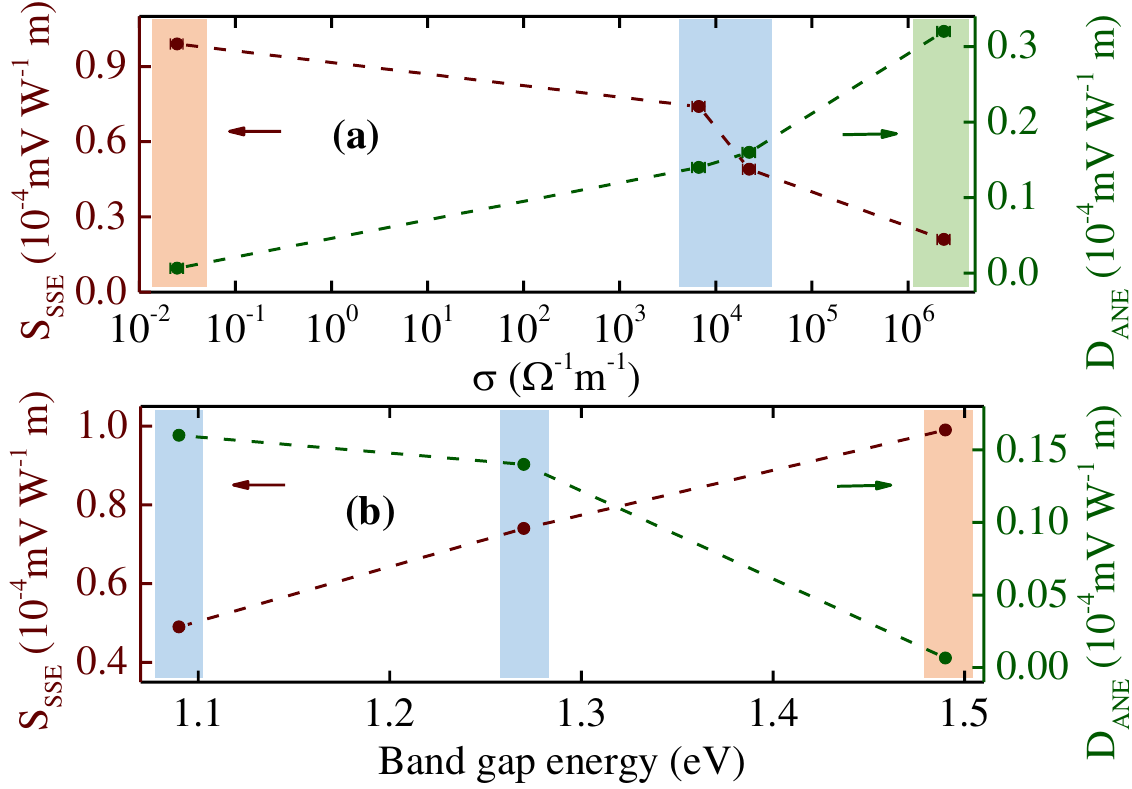}
    \caption{SSE and ANE coefficients as a function of (a) the electrical conductivity $\sigma$ for NiFe$_2$O$_{\textrm{x}_{\textrm{1}}/\textrm{x}_{\textrm{2}}}$ (blue area), NFO (orange area) and Ni$_{33}$Fe$_{67}$ (green area) samples and (b) the optical band gap for NiFe$_2$O$_{\textrm{x}_{\textrm{1}}/\textrm{x}_{\textrm{2}}}$ and NFO samples.}
    \label{fig:coefficients}
\end{figure}

Figure \ref{fig:coefficients}(a) shows the SSE ($S_{\textrm{SSE}}=\frac{Vsat_{\textrm{norm}}}{\phi_\textrm{q}}$) and ANE ($D_{\textrm{ANE}}=\frac{Vsat_{\textrm{norm}}}{\phi_\textrm{q}}$) coefficients extracted from the corresponding slopes of the curves in Fig. \ref{fig:slopes}, plotted against the RT value for the measured electrical conductivity. There is a pronounced increase of the $D_{\textrm{ANE}}$ when the conductivity increases, whereas the $S_{\textrm{SSE}}$ decreases.\par

Figure \ref{fig:coefficients}(b) depicts the dependence of the SSE and ANE coefficients on the optical band gap for the NFO and NiFe$_2$O$_{\textrm{x}_{\textrm{1}}/\textrm{x}_{\textrm{2}}}$ bilayers. A short description of the band gab determination can be found in the Supplemental Materials V \cite{Bou} (including Refs. \cite{Meinert:2014,Bougiatioti:2017,Lord:1960,Klewe:2014,Meier:2013}). It is clearly observed that the more conducting samples are characterized by lower band gap energies, reflecting the existence of additional electronic states in the band gap. Additionally, the ANE$^\textrm{intr}$ coefficient increases for decreasing band gap energy verifying the previous assumption of more mobile charge carriers at a reduced oxygen concentration. On the contrary, the SSE coefficient increases for larger band gap energies.

\begin{figure}[!ht]
    \centering
    \includegraphics[height=7cm, width=\linewidth]{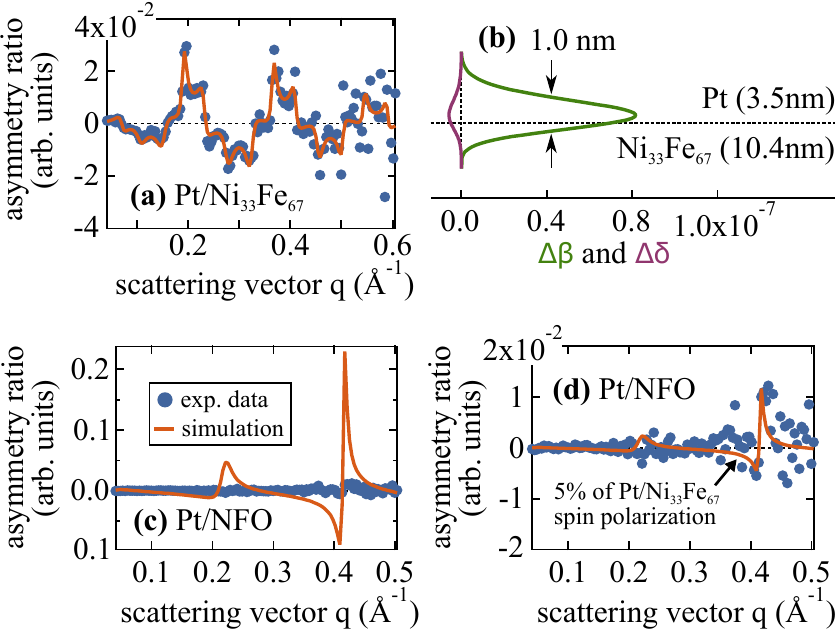}
  \caption{(a) XRMR asymmetry ratio for Pt/Ni$_{33}$Fe$_{67}$ and simulation with the corresponding magnetooptic depth profile (b). (c) XRMR asymmetry ratio for Pt/NFO after using the magnetooptic depth profile of (b), (d) assuming 5$\%$ of the Pt/Ni$_{33}$Fe$_{67}$ spin polarization.}
    \label{fig:XRMR}
\end{figure}

The absence of MPE in Pt/NFO, Pt/NiFe$_2$O$_{\textrm{x}_{\textrm{1}}/\textrm{x}_{\textrm{2}}}$ samples and the presence of MPE in the metallic Pt/Ni$_{33}$Fe$_{67}$ bilayer is now confirmed by XRMR (Fig. \ref{fig:XRMR}). In Fig. \ref{fig:XRMR}(a) the measured XRMR asymmetry ratio for the Pt/Ni$_{33}$Fe$_{67}$ bilayer is displayed. From the corresponding fitting and by comparing the experimental fit values of $\Delta\delta$ and $\Delta\beta$ derived from the magnetooptic depth profile in Fig. \ref{fig:XRMR}(b) to $\textit{ab initio}$ calculations \cite{Kuschel:2015}, we obtain a maximum Pt magnetic moment of $(0.48 \pm 0.08)\,\mu_{\textrm{B}}$ per spin polarized Pt atom, consistent with earlier results \cite{Klewe:2016}. The effective spin polarized Pt thickness is calculated to be $(1.0 \pm 0.1)\,$nm similar to our previous investigations \cite{Klewe:2016}.\par

In Fig. \ref{fig:XRMR}(c) the measured XRMR asymmetry ratio for the Pt/NFO bilayer is presented along with a simulation using a magnetooptic depth profile identical to the one derived for the Pt/Ni$_{33}$Fe$_{67}$ bilayer. Obviously, the simulated asymmetry ratio of the Pt/NFO sample (Fig. \ref{fig:XRMR}(c)) deviates strongly from the one of the Pt/Ni$_{33}$Fe$_{67}$ sample (Fig. \ref{fig:XRMR}(a)), although the same magnetooptic depth profile (Fig. \ref{fig:XRMR}(b)) was used. This is due to the different optical constants of Ni$_{33}$Fe$_{67}$ and NFO. Since no asymmetry was detected for the Pt/NFO sample, a potential MPE present in this film must be significantly smaller than in the all-metallic system.\par

By decreasing the magnitude of the magnetooptic parameters down to 5$\%$ of the Pt/Ni$_{33}$Fe$_{67}$ spin polarization (Fig. \ref{fig:XRMR}(d)), we can estimate a detection limit leading to an upper limit for the maximum magnetic moment in Pt of 0.04\,$\mu_{\textrm{B}}$ per spin polarized Pt atom. Moreover, for the Pt/NiFe$_2$O$_{\textrm{x}_{1}}$ and Pt/NiFe$_2$O$_{\textrm{x}_{2}}$ samples a detection limit of 0.1\,$\mu_{\textrm{B}}$ and 0.01\,$\mu_{\textrm{B}}$ per spin polarized Pt atom is extracted in the same way, see Supplemental Materials II \cite{Bou}. Finally, possible MPEs can be neglected down to these limits for all samples except for the metallic Pt/Ni$_{33}$Fe$_{67}$ bilayer, where a distinct spin polarization in the Pt layer can be observed.

In conclusion, we investigated thermal spin transport phenomena in Pt/FM bilayers and separated the intrinsic ANE in the FM and proximity-induced ANE contributions quantitatively from the LSSE for sputter-deposited NiFe$_2$O$_{\textrm{x}}$ bilayers. This new compact procedure is based on the preparation of twin samples (with and without Pt), different measurement geometries, the normalization to the heat flux instead of the thermal gradient, and the determination of important correction factors to obtain quantitative LSSE and ANE values. In our work, we extracted the dependence of the LSSE and intrinsic ANE coefficients on the band gap energy and on the electrical conductivity of the samples. Furthermore, possible static MPE in Pt were studied via XRMR. We found no magnetic response down to our detection limits of 0.04 $\mu_{\textrm{B}}$, 0.1 $\mu_{\textrm{B}}$ and 0.01 $\mu_{\textrm{B}}$ per spin polarized Pt atom for Pt/NFO, Pt/NiFe$_2$O$_{\textrm{x}_{1}}$ and Pt/NiFe$_2$O$_{\textrm{x}_{2}}$, respectively. For the Pt/Ni$_{33}$Fe$_{67}$ we calculated a maximum magnetic moment of 0.48 $\mu_{\textrm{B}}$ per spin polarized Pt atom. All XRMR results are well in line with the absence/presence of proximity-induced ANE contributions. In a next step this technique of thermal transport effect separation allows to study the individual transport effects depending on other properties of the samples, e.g., thicknesses and roughnesses. Even the proximity-induced thermal magnetotransport can be extracted experimentally as well as the LSSE can be identified in metallic films.\par

The authors gratefully acknowledge financial support by the Deutsche Forschungsgemeinschaft (DFG) within the Priority Program Spin Caloric Transport (SPP 1538) and the European Synchrotron Radiation Facility (ESRF). Christoph Klewe further acknowledges financial support by the Alexander von Humboldt foundation.



\bibliographystyle{apsrev4-1}
\bibliography{main}

\begin{thebibliography}{41}%
\makeatletter
\providecommand \@ifxundefined [1]{%
 \@ifx{#1\undefined}
}%
\providecommand \@ifnum [1]{%
 \ifnum #1\expandafter \@firstoftwo
 \else \expandafter \@secondoftwo
 \fi
}%
\providecommand \@ifx [1]{%
 \ifx #1\expandafter \@firstoftwo
 \else \expandafter \@secondoftwo
 \fi
}%
\providecommand \natexlab [1]{#1}%
\providecommand \enquote  [1]{``#1''}%
\providecommand \bibnamefont  [1]{#1}%
\providecommand \bibfnamefont [1]{#1}%
\providecommand \citenamefont [1]{#1}%
\providecommand \href@noop [0]{\@secondoftwo}%
\providecommand \href [0]{\begingroup \@sanitize@url \@href}%
\providecommand \@href[1]{\@@startlink{#1}\@@href}%
\providecommand \@@href[1]{\endgroup#1\@@endlink}%
\providecommand \@sanitize@url [0]{\catcode `\\12\catcode `\$12\catcode
  `\&12\catcode `\#12\catcode `\^12\catcode `\_12\catcode `\%12\relax}%
\providecommand \@@startlink[1]{}%
\providecommand \@@endlink[0]{}%
\providecommand \url  [0]{\begingroup\@sanitize@url \@url }%
\providecommand \@url [1]{\endgroup\@href {#1}{\urlprefix }}%
\providecommand \urlprefix  [0]{URL }%
\providecommand \Eprint [0]{\href }%
\providecommand \doibase [0]{http://dx.doi.org/}%
\providecommand \selectlanguage [0]{\@gobble}%
\providecommand \bibinfo  [0]{\@secondoftwo}%
\providecommand \bibfield  [0]{\@secondoftwo}%
\providecommand \translation [1]{[#1]}%
\providecommand \BibitemOpen [0]{}%
\providecommand \bibitemStop [0]{}%
\providecommand \bibitemNoStop [0]{.\EOS\space}%
\providecommand \EOS [0]{\spacefactor3000\relax}%
\providecommand \BibitemShut  [1]{\csname bibitem#1\endcsname}%
\let\auto@bib@innerbib\@empty
\bibitem [{\citenamefont {Hoffmann}\ and\ \citenamefont
  {Bader}(2015)}]{Hoffmann:2015}%
  \BibitemOpen
  \bibfield  {author} {\bibinfo {author} {\bibfnamefont {A.}~\bibnamefont
  {Hoffmann}}\ and\ \bibinfo {author} {\bibfnamefont {S.~D.}\ \bibnamefont
  {Bader}},\ }\href {\doibase 10.1103/PhysRevApplied.4.047001} {\bibfield
  {journal} {\bibinfo  {journal} {Phys. Rev. Applied}\ }\textbf {\bibinfo
  {volume} {4}},\ \bibinfo {pages} {047001} (\bibinfo {year}
  {2015})}\BibitemShut {NoStop}%
\bibitem [{\citenamefont {Bauer}\ \emph {et~al.}(2012)\citenamefont {Bauer},
  \citenamefont {Saitoh},\ and\ \citenamefont {van Wees}}]{Bauer:2012}%
  \BibitemOpen
  \bibfield  {author} {\bibinfo {author} {\bibfnamefont {G.~E.~W.}\
  \bibnamefont {Bauer}}, \bibinfo {author} {\bibfnamefont {E.}~\bibnamefont
  {Saitoh}}, \ and\ \bibinfo {author} {\bibfnamefont {B.~J.}\ \bibnamefont {van
  Wees}},\ }\href {\doibase 10.1038/nmat3301} {\bibfield  {journal} {\bibinfo
  {journal} {Nat. Mater.}\ }\textbf {\bibinfo {volume} {11}},\ \bibinfo {pages}
  {391} (\bibinfo {year} {2012})}\BibitemShut {NoStop}%
\bibitem [{\citenamefont {Hirsch}(1999)}]{Hirsch:1999}%
  \BibitemOpen
  \bibfield  {author} {\bibinfo {author} {\bibfnamefont {J.~E.}\ \bibnamefont
  {Hirsch}},\ }\href {\doibase 10.1103/PhysRevLett.83.1834} {\bibfield
  {journal} {\bibinfo  {journal} {Phys. Rev. Lett.}\ }\textbf {\bibinfo
  {volume} {83}},\ \bibinfo {pages} {1834} (\bibinfo {year}
  {1999})}\BibitemShut {NoStop}%
\bibitem [{\citenamefont {Uchida}\ \emph {et~al.}(2010)\citenamefont {Uchida},
  \citenamefont {Adachi}, \citenamefont {Ota}, \citenamefont {Nakayama},
  \citenamefont {Maekawa},\ and\ \citenamefont {Saitoh}}]{Uchida:2010}%
  \BibitemOpen
  \bibfield  {author} {\bibinfo {author} {\bibfnamefont {K.}~\bibnamefont
  {Uchida}}, \bibinfo {author} {\bibfnamefont {H.}~\bibnamefont {Adachi}},
  \bibinfo {author} {\bibfnamefont {T.}~\bibnamefont {Ota}}, \bibinfo {author}
  {\bibfnamefont {H.}~\bibnamefont {Nakayama}}, \bibinfo {author}
  {\bibfnamefont {S.}~\bibnamefont {Maekawa}}, \ and\ \bibinfo {author}
  {\bibfnamefont {E.}~\bibnamefont {Saitoh}},\ }\href
  {http://scitation.aip.org/content/aip/journal/apl/97/17/10.1063/1.3507386}
  {\bibfield  {journal} {\bibinfo  {journal} {Appl. Phys. Lett.}\ }\textbf
  {\bibinfo {volume} {97}},\ \bibinfo {eid} {172505} (\bibinfo {year}
  {2010})}\BibitemShut {NoStop}%
\bibitem [{\citenamefont {Uchida}\ \emph {et~al.}(2014)\citenamefont {Uchida},
  \citenamefont {Ishida}, \citenamefont {Kikkawa}, \citenamefont {Kirihara},
  \citenamefont {Murakami},\ and\ \citenamefont {Saitoh}}]{Uchida:2014}%
  \BibitemOpen
  \bibfield  {author} {\bibinfo {author} {\bibfnamefont {K.}~\bibnamefont
  {Uchida}}, \bibinfo {author} {\bibfnamefont {M.}~\bibnamefont {Ishida}},
  \bibinfo {author} {\bibfnamefont {T.}~\bibnamefont {Kikkawa}}, \bibinfo
  {author} {\bibfnamefont {A.}~\bibnamefont {Kirihara}}, \bibinfo {author}
  {\bibfnamefont {T.}~\bibnamefont {Murakami}}, \ and\ \bibinfo {author}
  {\bibfnamefont {E.}~\bibnamefont {Saitoh}},\ }\href
  {http://stacks.iop.org/0953-8984/26/i=34/a=343202} {\bibfield  {journal}
  {\bibinfo  {journal} {J. Phys.: Condens. Matter}\ }\textbf {\bibinfo {volume}
  {26}},\ \bibinfo {pages} {343202} (\bibinfo {year} {2014})}\BibitemShut
  {NoStop}%
\bibitem [{\citenamefont {Liu}\ \emph {et~al.}(2011)\citenamefont {Liu},
  \citenamefont {Moriyama}, \citenamefont {Ralph},\ and\ \citenamefont
  {Buhrman}}]{Liu:2011}%
  \BibitemOpen
  \bibfield  {author} {\bibinfo {author} {\bibfnamefont {L.}~\bibnamefont
  {Liu}}, \bibinfo {author} {\bibfnamefont {T.}~\bibnamefont {Moriyama}},
  \bibinfo {author} {\bibfnamefont {D.~C.}\ \bibnamefont {Ralph}}, \ and\
  \bibinfo {author} {\bibfnamefont {R.~A.}\ \bibnamefont {Buhrman}},\ }\href
  {\doibase 10.1103/PhysRevLett.106.036601} {\bibfield  {journal} {\bibinfo
  {journal} {Phys. Rev. Lett.}\ }\textbf {\bibinfo {volume} {106}},\ \bibinfo
  {pages} {036601} (\bibinfo {year} {2011})}\BibitemShut {NoStop}%
\bibitem [{\citenamefont {Liu}\ \emph {et~al.}(2012)\citenamefont {Liu},
  \citenamefont {Pai}, \citenamefont {Li}, \citenamefont {Tseng}, \citenamefont
  {Ralph},\ and\ \citenamefont {Buhrman}}]{Liu:2012}%
  \BibitemOpen
  \bibfield  {author} {\bibinfo {author} {\bibfnamefont {L.}~\bibnamefont
  {Liu}}, \bibinfo {author} {\bibfnamefont {C.-F.}\ \bibnamefont {Pai}},
  \bibinfo {author} {\bibfnamefont {Y.}~\bibnamefont {Li}}, \bibinfo {author}
  {\bibfnamefont {H.}~\bibnamefont {Tseng}}, \bibinfo {author} {\bibfnamefont
  {D.}~\bibnamefont {Ralph}}, \ and\ \bibinfo {author} {\bibfnamefont
  {R.}~\bibnamefont {Buhrman}},\ }\href@noop {} {\bibfield  {journal} {\bibinfo
   {journal} {Science}\ }\textbf {\bibinfo {volume} {336}},\ \bibinfo {pages}
  {555} (\bibinfo {year} {2012})}\BibitemShut {NoStop}%
\bibitem [{\citenamefont {Tang}\ \emph {et~al.}(2013)\citenamefont {Tang},
  \citenamefont {Kitamura}, \citenamefont {Shikoh}, \citenamefont {Ando},
  \citenamefont {Shinjo},\ and\ \citenamefont {Shiraishi}}]{Tang:2013}%
  \BibitemOpen
  \bibfield  {author} {\bibinfo {author} {\bibfnamefont {Z.}~\bibnamefont
  {Tang}}, \bibinfo {author} {\bibfnamefont {Y.}~\bibnamefont {Kitamura}},
  \bibinfo {author} {\bibfnamefont {E.}~\bibnamefont {Shikoh}}, \bibinfo
  {author} {\bibfnamefont {Y.}~\bibnamefont {Ando}}, \bibinfo {author}
  {\bibfnamefont {T.}~\bibnamefont {Shinjo}}, \ and\ \bibinfo {author}
  {\bibfnamefont {M.}~\bibnamefont {Shiraishi}},\ }\href
  {http://stacks.iop.org/1882-0786/6/i=8/a=083001} {\bibfield  {journal}
  {\bibinfo  {journal} {Appl. Phys. Express}\ }\textbf {\bibinfo {volume}
  {6}},\ \bibinfo {pages} {083001} (\bibinfo {year} {2013})}\BibitemShut
  {NoStop}%
\bibitem [{\citenamefont {Pai}\ \emph {et~al.}(2012)\citenamefont {Pai},
  \citenamefont {Liu}, \citenamefont {Li}, \citenamefont {Tseng}, \citenamefont
  {Ralph},\ and\ \citenamefont {Buhrman}}]{Pai:2012}%
  \BibitemOpen
  \bibfield  {author} {\bibinfo {author} {\bibfnamefont {C.-F.}\ \bibnamefont
  {Pai}}, \bibinfo {author} {\bibfnamefont {L.}~\bibnamefont {Liu}}, \bibinfo
  {author} {\bibfnamefont {Y.}~\bibnamefont {Li}}, \bibinfo {author}
  {\bibfnamefont {H.~W.}\ \bibnamefont {Tseng}}, \bibinfo {author}
  {\bibfnamefont {D.~C.}\ \bibnamefont {Ralph}}, \ and\ \bibinfo {author}
  {\bibfnamefont {R.~A.}\ \bibnamefont {Buhrman}},\ }\href
  {http://scitation.aip.org/content/aip/journal/apl/101/12/10.1063/1.4753947}
  {\bibfield  {journal} {\bibinfo  {journal} {Appl. Phys. Lett.}\ }\textbf
  {\bibinfo {volume} {101}},\ \bibinfo {eid} {122404} (\bibinfo {year}
  {2012})}\BibitemShut {NoStop}%
\bibitem [{\citenamefont {Saitoh}\ \emph {et~al.}(2006)\citenamefont {Saitoh},
  \citenamefont {Ueda}, \citenamefont {Miyajima},\ and\ \citenamefont
  {Tatara}}]{Saitoh:2006}%
  \BibitemOpen
  \bibfield  {author} {\bibinfo {author} {\bibfnamefont {E.}~\bibnamefont
  {Saitoh}}, \bibinfo {author} {\bibfnamefont {M.}~\bibnamefont {Ueda}},
  \bibinfo {author} {\bibfnamefont {H.}~\bibnamefont {Miyajima}}, \ and\
  \bibinfo {author} {\bibfnamefont {G.}~\bibnamefont {Tatara}},\ }\href
  {http://scitation.aip.org/content/aip/journal/apl/88/18/10.1063/1.2199473}
  {\bibfield  {journal} {\bibinfo  {journal} {Appl. Phys. Lett.}\ }\textbf
  {\bibinfo {volume} {88}},\ \bibinfo {eid} {182509} (\bibinfo {year}
  {2006})}\BibitemShut {NoStop}%
\bibitem [{\citenamefont {Stoner}(1938)}]{Stoner:1938}%
  \BibitemOpen
  \bibfield  {author} {\bibinfo {author} {\bibfnamefont {E.~C.}\ \bibnamefont
  {Stoner}},\ }\href {\doibase 10.1098/rspa.1938.0066} {\bibfield  {journal}
  {\bibinfo  {journal} {Proc. R. Soc. London Ser. A}\ }\textbf {\bibinfo
  {volume} {165}},\ \bibinfo {pages} {372} (\bibinfo {year}
  {1938})}\BibitemShut {NoStop}%
\bibitem [{\citenamefont {Huang}\ \emph {et~al.}(2012)\citenamefont {Huang},
  \citenamefont {Fan}, \citenamefont {Qu}, \citenamefont {Chen}, \citenamefont
  {Wang}, \citenamefont {Wu}, \citenamefont {Chen}, \citenamefont {Xiao},\ and\
  \citenamefont {Chien}}]{Huang:2012}%
  \BibitemOpen
  \bibfield  {author} {\bibinfo {author} {\bibfnamefont {S.~Y.}\ \bibnamefont
  {Huang}}, \bibinfo {author} {\bibfnamefont {X.}~\bibnamefont {Fan}}, \bibinfo
  {author} {\bibfnamefont {D.}~\bibnamefont {Qu}}, \bibinfo {author}
  {\bibfnamefont {Y.~P.}\ \bibnamefont {Chen}}, \bibinfo {author}
  {\bibfnamefont {W.~G.}\ \bibnamefont {Wang}}, \bibinfo {author}
  {\bibfnamefont {J.}~\bibnamefont {Wu}}, \bibinfo {author} {\bibfnamefont
  {T.~Y.}\ \bibnamefont {Chen}}, \bibinfo {author} {\bibfnamefont {J.~Q.}\
  \bibnamefont {Xiao}}, \ and\ \bibinfo {author} {\bibfnamefont {C.~L.}\
  \bibnamefont {Chien}},\ }\href {\doibase 10.1103/PhysRevLett.109.107204}
  {\bibfield  {journal} {\bibinfo  {journal} {Phys. Rev. Lett.}\ }\textbf
  {\bibinfo {volume} {109}},\ \bibinfo {pages} {107204} (\bibinfo {year}
  {2012})}\BibitemShut {NoStop}%
\bibitem [{\citenamefont {Meier}\ \emph {et~al.}(2013)\citenamefont {Meier},
  \citenamefont {Kuschel}, \citenamefont {Shen}, \citenamefont {Gupta},
  \citenamefont {Kikkawa}, \citenamefont {Uchida}, \citenamefont {Saitoh},
  \citenamefont {Schmalhorst},\ and\ \citenamefont {Reiss}}]{Meier:2013}%
  \BibitemOpen
  \bibfield  {author} {\bibinfo {author} {\bibfnamefont {D.}~\bibnamefont
  {Meier}}, \bibinfo {author} {\bibfnamefont {T.}~\bibnamefont {Kuschel}},
  \bibinfo {author} {\bibfnamefont {L.}~\bibnamefont {Shen}}, \bibinfo {author}
  {\bibfnamefont {A.}~\bibnamefont {Gupta}}, \bibinfo {author} {\bibfnamefont
  {T.}~\bibnamefont {Kikkawa}}, \bibinfo {author} {\bibfnamefont
  {K.}~\bibnamefont {Uchida}}, \bibinfo {author} {\bibfnamefont
  {E.}~\bibnamefont {Saitoh}}, \bibinfo {author} {\bibfnamefont {J.-M.}\
  \bibnamefont {Schmalhorst}}, \ and\ \bibinfo {author} {\bibfnamefont
  {G.}~\bibnamefont {Reiss}},\ }\href {\doibase 10.1103/PhysRevB.87.054421}
  {\bibfield  {journal} {\bibinfo  {journal} {Phys. Rev. B}\ }\textbf {\bibinfo
  {volume} {87}},\ \bibinfo {pages} {054421} (\bibinfo {year}
  {2013})}\BibitemShut {NoStop}%
\bibitem [{\citenamefont {Ramos}\ \emph {et~al.}(2013)\citenamefont {Ramos},
  \citenamefont {Kikkawa}, \citenamefont {Uchida}, \citenamefont {Adachi},
  \citenamefont {Lucas}, \citenamefont {Aguirre}, \citenamefont {Algarabel},
  \citenamefont {Morellón}, \citenamefont {Maekawa}, \citenamefont {Saitoh},\
  and\ \citenamefont {Ibarra}}]{Ramos:2013}%
  \BibitemOpen
  \bibfield  {author} {\bibinfo {author} {\bibfnamefont {R.}~\bibnamefont
  {Ramos}}, \bibinfo {author} {\bibfnamefont {T.}~\bibnamefont {Kikkawa}},
  \bibinfo {author} {\bibfnamefont {K.}~\bibnamefont {Uchida}}, \bibinfo
  {author} {\bibfnamefont {H.}~\bibnamefont {Adachi}}, \bibinfo {author}
  {\bibfnamefont {I.}~\bibnamefont {Lucas}}, \bibinfo {author} {\bibfnamefont
  {M.~H.}\ \bibnamefont {Aguirre}}, \bibinfo {author} {\bibfnamefont
  {P.}~\bibnamefont {Algarabel}}, \bibinfo {author} {\bibfnamefont
  {L.}~\bibnamefont {Morellón}}, \bibinfo {author} {\bibfnamefont
  {S.}~\bibnamefont {Maekawa}}, \bibinfo {author} {\bibfnamefont
  {E.}~\bibnamefont {Saitoh}}, \ and\ \bibinfo {author} {\bibfnamefont {M.~R.}\
  \bibnamefont {Ibarra}},\ }\href
  {http://scitation.aip.org/content/aip/journal/apl/102/7/10.1063/1.4793486}
  {\bibfield  {journal} {\bibinfo  {journal} {Appl. Phys. Lett.}\ }\textbf
  {\bibinfo {volume} {102}},\ \bibinfo {eid} {072413} (\bibinfo {year}
  {2013})}\BibitemShut {NoStop}%
\bibitem [{\citenamefont {Ramos}\ \emph {et~al.}(2014)\citenamefont {Ramos},
  \citenamefont {Aguirre}, \citenamefont {Anad\'on}, \citenamefont {Blasco},
  \citenamefont {Lucas}, \citenamefont {Uchida}, \citenamefont {Algarabel},
  \citenamefont {Morell\'on}, \citenamefont {Saitoh},\ and\ \citenamefont
  {Ibarra}}]{Ramos:2014}%
  \BibitemOpen
  \bibfield  {author} {\bibinfo {author} {\bibfnamefont {R.}~\bibnamefont
  {Ramos}}, \bibinfo {author} {\bibfnamefont {M.~H.}\ \bibnamefont {Aguirre}},
  \bibinfo {author} {\bibfnamefont {A.}~\bibnamefont {Anad\'on}}, \bibinfo
  {author} {\bibfnamefont {J.}~\bibnamefont {Blasco}}, \bibinfo {author}
  {\bibfnamefont {I.}~\bibnamefont {Lucas}}, \bibinfo {author} {\bibfnamefont
  {K.}~\bibnamefont {Uchida}}, \bibinfo {author} {\bibfnamefont {P.~A.}\
  \bibnamefont {Algarabel}}, \bibinfo {author} {\bibfnamefont {L.}~\bibnamefont
  {Morell\'on}}, \bibinfo {author} {\bibfnamefont {E.}~\bibnamefont {Saitoh}},
  \ and\ \bibinfo {author} {\bibfnamefont {M.~R.}\ \bibnamefont {Ibarra}},\
  }\href {\doibase 10.1103/PhysRevB.90.054422} {\bibfield  {journal} {\bibinfo
  {journal} {Phys. Rev. B}\ }\textbf {\bibinfo {volume} {90}},\ \bibinfo
  {pages} {054422} (\bibinfo {year} {2014})}\BibitemShut {NoStop}%
\bibitem [{\citenamefont {Ramos}\ \emph {et~al.}(2015)\citenamefont {Ramos},
  \citenamefont {Kikkawa}, \citenamefont {Aguirre}, \citenamefont {Lucas},
  \citenamefont {Anad\'on}, \citenamefont {Oyake}, \citenamefont {Uchida},
  \citenamefont {Adachi}, \citenamefont {Shiomi}, \citenamefont {Algarabel},
  \citenamefont {Morell\'on}, \citenamefont {Maekawa}, \citenamefont {Saitoh},\
  and\ \citenamefont {Ibarra}}]{Ramos:2015}%
  \BibitemOpen
  \bibfield  {author} {\bibinfo {author} {\bibfnamefont {R.}~\bibnamefont
  {Ramos}}, \bibinfo {author} {\bibfnamefont {T.}~\bibnamefont {Kikkawa}},
  \bibinfo {author} {\bibfnamefont {M.~H.}\ \bibnamefont {Aguirre}}, \bibinfo
  {author} {\bibfnamefont {I.}~\bibnamefont {Lucas}}, \bibinfo {author}
  {\bibfnamefont {A.}~\bibnamefont {Anad\'on}}, \bibinfo {author}
  {\bibfnamefont {T.}~\bibnamefont {Oyake}}, \bibinfo {author} {\bibfnamefont
  {K.}~\bibnamefont {Uchida}}, \bibinfo {author} {\bibfnamefont
  {H.}~\bibnamefont {Adachi}}, \bibinfo {author} {\bibfnamefont
  {J.}~\bibnamefont {Shiomi}}, \bibinfo {author} {\bibfnamefont {P.~A.}\
  \bibnamefont {Algarabel}}, \bibinfo {author} {\bibfnamefont {L.}~\bibnamefont
  {Morell\'on}}, \bibinfo {author} {\bibfnamefont {S.}~\bibnamefont {Maekawa}},
  \bibinfo {author} {\bibfnamefont {E.}~\bibnamefont {Saitoh}}, \ and\ \bibinfo
  {author} {\bibfnamefont {M.~R.}\ \bibnamefont {Ibarra}},\ }\href {\doibase
  10.1103/PhysRevB.92.220407} {\bibfield  {journal} {\bibinfo  {journal} {Phys.
  Rev. B}\ }\textbf {\bibinfo {volume} {92}},\ \bibinfo {pages} {220407}
  (\bibinfo {year} {2015})}\BibitemShut {NoStop}%
\bibitem [{\citenamefont {Ramos}\ \emph {et~al.}(2016)\citenamefont {Ramos},
  \citenamefont {Anadón}, \citenamefont {Lucas}, \citenamefont {Uchida},
  \citenamefont {Algarabel}, \citenamefont {Morellón}, \citenamefont
  {Aguirre}, \citenamefont {Saitoh},\ and\ \citenamefont
  {Ibarra}}]{Ramos:2016}%
  \BibitemOpen
  \bibfield  {author} {\bibinfo {author} {\bibfnamefont {R.}~\bibnamefont
  {Ramos}}, \bibinfo {author} {\bibfnamefont {A.}~\bibnamefont {Anadón}},
  \bibinfo {author} {\bibfnamefont {I.}~\bibnamefont {Lucas}}, \bibinfo
  {author} {\bibfnamefont {K.}~\bibnamefont {Uchida}}, \bibinfo {author}
  {\bibfnamefont {P.~A.}\ \bibnamefont {Algarabel}}, \bibinfo {author}
  {\bibfnamefont {L.}~\bibnamefont {Morellón}}, \bibinfo {author}
  {\bibfnamefont {M.~H.}\ \bibnamefont {Aguirre}}, \bibinfo {author}
  {\bibfnamefont {E.}~\bibnamefont {Saitoh}}, \ and\ \bibinfo {author}
  {\bibfnamefont {M.~R.}\ \bibnamefont {Ibarra}},\ }\href {\doibase
  10.1063/1.4950994} {\bibfield  {journal} {\bibinfo  {journal} {APL Mater.}\
  }\textbf {\bibinfo {volume} {4}},\ \bibinfo {pages} {104802} (\bibinfo {year}
  {2016})}\BibitemShut {NoStop}%
\bibitem [{\citenamefont {Wu}\ \emph {et~al.}(2014)\citenamefont {Wu},
  \citenamefont {Hoffman}, \citenamefont {Pearson},\ and\ \citenamefont
  {Bhattacharya}}]{Wu:2014}%
  \BibitemOpen
  \bibfield  {author} {\bibinfo {author} {\bibfnamefont {S.~M.}\ \bibnamefont
  {Wu}}, \bibinfo {author} {\bibfnamefont {J.}~\bibnamefont {Hoffman}},
  \bibinfo {author} {\bibfnamefont {J.~E.}\ \bibnamefont {Pearson}}, \ and\
  \bibinfo {author} {\bibfnamefont {A.}~\bibnamefont {Bhattacharya}},\
  }\href@noop {} {\bibfield  {journal} {\bibinfo  {journal} {Appl. Phys.
  Lett.}\ }\textbf {\bibinfo {volume} {105}},\ \bibinfo {pages} {092409}
  (\bibinfo {year} {2014})}\BibitemShut {NoStop}%
\bibitem [{\citenamefont {Lee}\ \emph {et~al.}(2015)\citenamefont {Lee},
  \citenamefont {Kim}, \citenamefont {Lee}, \citenamefont {Kim}, \citenamefont
  {Lee}, \citenamefont {Lee}, \citenamefont {Jeong}, \citenamefont {Lee},
  \citenamefont {Song}, \citenamefont {Sohn}, \citenamefont {Shin},\ and\
  \citenamefont {Park}}]{Lee:2015}%
  \BibitemOpen
  \bibfield  {author} {\bibinfo {author} {\bibfnamefont {K.-D.}\ \bibnamefont
  {Lee}}, \bibinfo {author} {\bibfnamefont {D.-J.}\ \bibnamefont {Kim}},
  \bibinfo {author} {\bibfnamefont {H.~Y.}\ \bibnamefont {Lee}}, \bibinfo
  {author} {\bibfnamefont {S.-H.}\ \bibnamefont {Kim}}, \bibinfo {author}
  {\bibfnamefont {J.-H.}\ \bibnamefont {Lee}}, \bibinfo {author} {\bibfnamefont
  {K.-M.}\ \bibnamefont {Lee}}, \bibinfo {author} {\bibfnamefont {J.-R.}\
  \bibnamefont {Jeong}}, \bibinfo {author} {\bibfnamefont {K.-S.}\ \bibnamefont
  {Lee}}, \bibinfo {author} {\bibfnamefont {H.-S.}\ \bibnamefont {Song}},
  \bibinfo {author} {\bibfnamefont {J.-W.}\ \bibnamefont {Sohn}}, \bibinfo
  {author} {\bibfnamefont {S.-C.}\ \bibnamefont {Shin}}, \ and\ \bibinfo
  {author} {\bibfnamefont {B.-G.}\ \bibnamefont {Park}},\ }\href@noop {}
  {\bibfield  {journal} {\bibinfo  {journal} {Sci. Rep.}\ }\textbf {\bibinfo
  {volume} {5}},\ \bibinfo {pages} {10249} (\bibinfo {year}
  {2015})}\BibitemShut {NoStop}%
\bibitem [{\citenamefont {Uchida}\ \emph {et~al.}(2015)\citenamefont {Uchida},
  \citenamefont {Kikkawa}, \citenamefont {Seki}, \citenamefont {Oyake},
  \citenamefont {Shiomi}, \citenamefont {Qiu}, \citenamefont {Takanashi},\ and\
  \citenamefont {Saitoh}}]{Uchida:2015}%
  \BibitemOpen
  \bibfield  {author} {\bibinfo {author} {\bibfnamefont {K.}~\bibnamefont
  {Uchida}}, \bibinfo {author} {\bibfnamefont {T.}~\bibnamefont {Kikkawa}},
  \bibinfo {author} {\bibfnamefont {T.}~\bibnamefont {Seki}}, \bibinfo {author}
  {\bibfnamefont {T.}~\bibnamefont {Oyake}}, \bibinfo {author} {\bibfnamefont
  {J.}~\bibnamefont {Shiomi}}, \bibinfo {author} {\bibfnamefont
  {Z.}~\bibnamefont {Qiu}}, \bibinfo {author} {\bibfnamefont {K.}~\bibnamefont
  {Takanashi}}, \ and\ \bibinfo {author} {\bibfnamefont {E.}~\bibnamefont
  {Saitoh}},\ }\href {\doibase 10.1103/PhysRevB.92.094414} {\bibfield
  {journal} {\bibinfo  {journal} {Phys. Rev. B}\ }\textbf {\bibinfo {volume}
  {92}},\ \bibinfo {pages} {094414} (\bibinfo {year} {2015})}\BibitemShut
  {NoStop}%
\bibitem [{\citenamefont {Uchida}\ \emph {et~al.}(2016)\citenamefont {Uchida},
  \citenamefont {Adachi}, \citenamefont {Kikkawa}, \citenamefont {Kirihara},
  \citenamefont {Ishida}, \citenamefont {Yorozu}, \citenamefont {Maekawa},\
  and\ \citenamefont {Saitoh}}]{Uchida:2016}%
  \BibitemOpen
  \bibfield  {author} {\bibinfo {author} {\bibfnamefont {K.}~\bibnamefont
  {Uchida}}, \bibinfo {author} {\bibfnamefont {H.}~\bibnamefont {Adachi}},
  \bibinfo {author} {\bibfnamefont {T.}~\bibnamefont {Kikkawa}}, \bibinfo
  {author} {\bibfnamefont {A.}~\bibnamefont {Kirihara}}, \bibinfo {author}
  {\bibfnamefont {M.}~\bibnamefont {Ishida}}, \bibinfo {author} {\bibfnamefont
  {S.}~\bibnamefont {Yorozu}}, \bibinfo {author} {\bibfnamefont
  {S.}~\bibnamefont {Maekawa}}, \ and\ \bibinfo {author} {\bibfnamefont
  {E.}~\bibnamefont {Saitoh}},\ }\href@noop {} {\bibfield  {journal} {\bibinfo
  {journal} {Proc. IEEE}\ }\textbf {\bibinfo {volume} {104}},\ \bibinfo {pages}
  {1946} (\bibinfo {year} {2016})}\BibitemShut {NoStop}%
\bibitem [{FOO()}]{FOOTNOTE}%
  \BibitemOpen
  \href@noop {} {\bibinfo  {journal} {Bauer \textit{et al}. \cite{Bauer:2012}
  introduced the term ``spin Seebeck effect" for FMI systems while for the
  spin-dependent thermal creation of moving electrons in metals and
  semiconductors they suggested the term ``spin-dependent Seebeck effect"
  (SDSE). In our letter we will not distinguish between SSE and SDSE in the
  conducting materials and, thus, we just use the term SSE throughout the
  letter}\ }\BibitemShut {NoStop}%
\bibitem [{\citenamefont {Kikkawa}\ \emph
  {et~al.}(2013{\natexlab{a}})\citenamefont {Kikkawa}, \citenamefont {Uchida},
  \citenamefont {Shiomi}, \citenamefont {Qiu}, \citenamefont {Hou},
  \citenamefont {Tian}, \citenamefont {Nakayama}, \citenamefont {Jin},\ and\
  \citenamefont {Saitoh}}]{Kikkawa:2013PRL}%
  \BibitemOpen
\bibfield  {journal} {  }\bibfield  {author} {\bibinfo {author} {\bibfnamefont
  {T.}~\bibnamefont {Kikkawa}}, \bibinfo {author} {\bibfnamefont
  {K.}~\bibnamefont {Uchida}}, \bibinfo {author} {\bibfnamefont
  {Y.}~\bibnamefont {Shiomi}}, \bibinfo {author} {\bibfnamefont
  {Z.}~\bibnamefont {Qiu}}, \bibinfo {author} {\bibfnamefont {D.}~\bibnamefont
  {Hou}}, \bibinfo {author} {\bibfnamefont {D.}~\bibnamefont {Tian}}, \bibinfo
  {author} {\bibfnamefont {H.}~\bibnamefont {Nakayama}}, \bibinfo {author}
  {\bibfnamefont {X.-F.}\ \bibnamefont {Jin}}, \ and\ \bibinfo {author}
  {\bibfnamefont {E.}~\bibnamefont {Saitoh}},\ }\href {\doibase
  10.1103/PhysRevLett.110.067207} {\bibfield  {journal} {\bibinfo  {journal}
  {Phys. Rev. Lett.}\ }\textbf {\bibinfo {volume} {110}},\ \bibinfo {pages}
  {067207} (\bibinfo {year} {2013}{\natexlab{a}})}\BibitemShut {NoStop}%
\bibitem [{\citenamefont {Xu}\ \emph {et~al.}(2014)\citenamefont {Xu},
  \citenamefont {Yang}, \citenamefont {Tang}, \citenamefont {Jiang},
  \citenamefont {Schneider}, \citenamefont {Whig},\ and\ \citenamefont
  {Shi}}]{Xu:2014}%
  \BibitemOpen
  \bibfield  {author} {\bibinfo {author} {\bibfnamefont {Y.}~\bibnamefont
  {Xu}}, \bibinfo {author} {\bibfnamefont {B.}~\bibnamefont {Yang}}, \bibinfo
  {author} {\bibfnamefont {C.}~\bibnamefont {Tang}}, \bibinfo {author}
  {\bibfnamefont {Z.}~\bibnamefont {Jiang}}, \bibinfo {author} {\bibfnamefont
  {M.}~\bibnamefont {Schneider}}, \bibinfo {author} {\bibfnamefont
  {R.}~\bibnamefont {Whig}}, \ and\ \bibinfo {author} {\bibfnamefont
  {J.}~\bibnamefont {Shi}},\ }\href {\doibase 10.1063/1.4904467} {\bibfield
  {journal} {\bibinfo  {journal} {Appl. Phys. Lett.}\ }\textbf {\bibinfo
  {volume} {105}},\ \bibinfo {pages} {242404} (\bibinfo {year}
  {2014})}\BibitemShut {NoStop}%
\bibitem [{\citenamefont {Miao}\ \emph {et~al.}(2016)\citenamefont {Miao},
  \citenamefont {Huang}, \citenamefont {Qu},\ and\ \citenamefont
  {Chien}}]{Miao:2016}%
  \BibitemOpen
  \bibfield  {author} {\bibinfo {author} {\bibfnamefont {B.~F.}\ \bibnamefont
  {Miao}}, \bibinfo {author} {\bibfnamefont {S.~Y.}\ \bibnamefont {Huang}},
  \bibinfo {author} {\bibfnamefont {D.}~\bibnamefont {Qu}}, \ and\ \bibinfo
  {author} {\bibfnamefont {C.~L.}\ \bibnamefont {Chien}},\ }\href {\doibase
  10.1063/1.4941340} {\bibfield  {journal} {\bibinfo  {journal} {AIP Adv.}\
  }\textbf {\bibinfo {volume} {6}},\ \bibinfo {pages} {015018} (\bibinfo {year}
  {2016})}\BibitemShut {NoStop}%
\bibitem [{\citenamefont {Kikkawa}\ \emph
  {et~al.}(2013{\natexlab{b}})\citenamefont {Kikkawa}, \citenamefont {Uchida},
  \citenamefont {Daimon}, \citenamefont {Shiomi}, \citenamefont {Adachi},
  \citenamefont {Qiu}, \citenamefont {Hou}, \citenamefont {Jin}, \citenamefont
  {Maekawa},\ and\ \citenamefont {Saitoh}}]{Kikkawa:2013PRB}%
  \BibitemOpen
  \bibfield  {author} {\bibinfo {author} {\bibfnamefont {T.}~\bibnamefont
  {Kikkawa}}, \bibinfo {author} {\bibfnamefont {K.}~\bibnamefont {Uchida}},
  \bibinfo {author} {\bibfnamefont {S.}~\bibnamefont {Daimon}}, \bibinfo
  {author} {\bibfnamefont {Y.}~\bibnamefont {Shiomi}}, \bibinfo {author}
  {\bibfnamefont {H.}~\bibnamefont {Adachi}}, \bibinfo {author} {\bibfnamefont
  {Z.}~\bibnamefont {Qiu}}, \bibinfo {author} {\bibfnamefont {D.}~\bibnamefont
  {Hou}}, \bibinfo {author} {\bibfnamefont {X.-F.}\ \bibnamefont {Jin}},
  \bibinfo {author} {\bibfnamefont {S.}~\bibnamefont {Maekawa}}, \ and\
  \bibinfo {author} {\bibfnamefont {E.}~\bibnamefont {Saitoh}},\ }\href
  {\doibase 10.1103/PhysRevB.88.214403} {\bibfield  {journal} {\bibinfo
  {journal} {Phys. Rev. B}\ }\textbf {\bibinfo {volume} {88}},\ \bibinfo
  {pages} {214403} (\bibinfo {year} {2013}{\natexlab{b}})}\BibitemShut
  {NoStop}%
\bibitem [{\citenamefont {Macke}\ and\ \citenamefont
  {Goering}(2014)}]{Macke:2014}%
  \BibitemOpen
  \bibfield  {author} {\bibinfo {author} {\bibfnamefont {S.}~\bibnamefont
  {Macke}}\ and\ \bibinfo {author} {\bibfnamefont {E.}~\bibnamefont
  {Goering}},\ }\href {http://stacks.iop.org/0953-8984/26/i=36/a=363201}
  {\bibfield  {journal} {\bibinfo  {journal} {J. Phys.: Condens. Matter}\
  }\textbf {\bibinfo {volume} {26}},\ \bibinfo {pages} {363201} (\bibinfo
  {year} {2014})}\BibitemShut {NoStop}%
\bibitem [{\citenamefont {Kuschel}\ \emph {et~al.}(2015)\citenamefont
  {Kuschel}, \citenamefont {Klewe}, \citenamefont {Schmalhorst}, \citenamefont
  {Bertram}, \citenamefont {Kuschel}, \citenamefont {Schemme}, \citenamefont
  {Wollschl\"ager}, \citenamefont {Francoual}, \citenamefont {Strempfer},
  \citenamefont {Gupta}, \citenamefont {Meinert}, \citenamefont {G\"otz},
  \citenamefont {Meier},\ and\ \citenamefont {Reiss}}]{Kuschel:2015}%
  \BibitemOpen
  \bibfield  {author} {\bibinfo {author} {\bibfnamefont {T.}~\bibnamefont
  {Kuschel}}, \bibinfo {author} {\bibfnamefont {C.}~\bibnamefont {Klewe}},
  \bibinfo {author} {\bibfnamefont {J.-M.}\ \bibnamefont {Schmalhorst}},
  \bibinfo {author} {\bibfnamefont {F.}~\bibnamefont {Bertram}}, \bibinfo
  {author} {\bibfnamefont {O.}~\bibnamefont {Kuschel}}, \bibinfo {author}
  {\bibfnamefont {T.}~\bibnamefont {Schemme}}, \bibinfo {author} {\bibfnamefont
  {J.}~\bibnamefont {Wollschl\"ager}}, \bibinfo {author} {\bibfnamefont
  {S.}~\bibnamefont {Francoual}}, \bibinfo {author} {\bibfnamefont
  {J.}~\bibnamefont {Strempfer}}, \bibinfo {author} {\bibfnamefont
  {A.}~\bibnamefont {Gupta}}, \bibinfo {author} {\bibfnamefont
  {M.}~\bibnamefont {Meinert}}, \bibinfo {author} {\bibfnamefont
  {G.}~\bibnamefont {G\"otz}}, \bibinfo {author} {\bibfnamefont
  {D.}~\bibnamefont {Meier}}, \ and\ \bibinfo {author} {\bibfnamefont
  {G.}~\bibnamefont {Reiss}},\ }\href {\doibase 10.1103/PhysRevLett.115.097401}
  {\bibfield  {journal} {\bibinfo  {journal} {Phys. Rev. Lett.}\ }\textbf
  {\bibinfo {volume} {115}},\ \bibinfo {pages} {097401} (\bibinfo {year}
  {2015})}\BibitemShut {NoStop}%
\bibitem [{\citenamefont {Brown}\ \emph {et~al.}(2001)\citenamefont {Brown},
  \citenamefont {Bouchenoire}, \citenamefont {Bowyer}, \citenamefont {Kervin},
  \citenamefont {Laundy}, \citenamefont {Longfield}, \citenamefont {Mannix},
  \citenamefont {Paul}, \citenamefont {Stunault}, \citenamefont {Thompson},
  \citenamefont {Cooper}, \citenamefont {Lucas},\ and\ \citenamefont
  {Stirling}}]{Brown:2001}%
  \BibitemOpen
  \bibfield  {author} {\bibinfo {author} {\bibfnamefont {S.~D.}\ \bibnamefont
  {Brown}}, \bibinfo {author} {\bibfnamefont {L.}~\bibnamefont {Bouchenoire}},
  \bibinfo {author} {\bibfnamefont {D.}~\bibnamefont {Bowyer}}, \bibinfo
  {author} {\bibfnamefont {J.}~\bibnamefont {Kervin}}, \bibinfo {author}
  {\bibfnamefont {D.}~\bibnamefont {Laundy}}, \bibinfo {author} {\bibfnamefont
  {M.~J.}\ \bibnamefont {Longfield}}, \bibinfo {author} {\bibfnamefont
  {D.}~\bibnamefont {Mannix}}, \bibinfo {author} {\bibfnamefont {D.~F.}\
  \bibnamefont {Paul}}, \bibinfo {author} {\bibfnamefont {A.}~\bibnamefont
  {Stunault}}, \bibinfo {author} {\bibfnamefont {P.}~\bibnamefont {Thompson}},
  \bibinfo {author} {\bibfnamefont {M.~J.}\ \bibnamefont {Cooper}}, \bibinfo
  {author} {\bibfnamefont {C.~A.}\ \bibnamefont {Lucas}}, \ and\ \bibinfo
  {author} {\bibfnamefont {W.~G.}\ \bibnamefont {Stirling}},\ }\href {\doibase
  10.1107/S0909049501015394} {\bibfield  {journal} {\bibinfo  {journal} {J.
  Synchrotron Radiat.}\ }\textbf {\bibinfo {volume} {8}},\ \bibinfo {pages}
  {1172} (\bibinfo {year} {2001})}\BibitemShut {NoStop}%
\bibitem [{\citenamefont {Bougiatioti}\ \emph {et~al.}()\citenamefont
  {Bougiatioti}, \citenamefont {Klewe}, \citenamefont {Meier}, \citenamefont
  {Manos}, \citenamefont {Kuschel}, \citenamefont {Wollschläger},
  \citenamefont {Bouchenoire}, \citenamefont {Brown}, \citenamefont
  {Schmalhorst}, \citenamefont {Reiss},\ and\ \citenamefont {Kuschel}}]{Bou}%
  \BibitemOpen
  \bibfield  {author} {\bibinfo {author} {\bibfnamefont {P.}~\bibnamefont
  {Bougiatioti}}, \bibinfo {author} {\bibfnamefont {C.}~\bibnamefont {Klewe}},
  \bibinfo {author} {\bibfnamefont {D.}~\bibnamefont {Meier}}, \bibinfo
  {author} {\bibfnamefont {O.}~\bibnamefont {Manos}}, \bibinfo {author}
  {\bibfnamefont {O.}~\bibnamefont {Kuschel}}, \bibinfo {author} {\bibfnamefont
  {J.}~\bibnamefont {Wollschläger}}, \bibinfo {author} {\bibfnamefont
  {L.}~\bibnamefont {Bouchenoire}}, \bibinfo {author} {\bibfnamefont {S.~D.}\
  \bibnamefont {Brown}}, \bibinfo {author} {\bibfnamefont {J.-M.}\ \bibnamefont
  {Schmalhorst}}, \bibinfo {author} {\bibfnamefont {G.}~\bibnamefont {Reiss}},
  \ and\ \bibinfo {author} {\bibfnamefont {T.}~\bibnamefont {Kuschel}},\
  }\href@noop {} {\bibinfo  {journal} {Supplemental Materials}\ }\BibitemShut
  {NoStop}%
\bibitem [{\citenamefont {Bouchenoire}\ \emph {et~al.}(2003)\citenamefont
  {Bouchenoire}, \citenamefont {Brown}, \citenamefont {Thompson}, \citenamefont
  {Duffy}, \citenamefont {Taylor},\ and\ \citenamefont
  {Cooper}}]{Bouchenoire:2003}%
  \BibitemOpen
\bibfield  {journal} {  }\bibfield  {author} {\bibinfo {author} {\bibfnamefont
  {L.}~\bibnamefont {Bouchenoire}}, \bibinfo {author} {\bibfnamefont {S.~D.}\
  \bibnamefont {Brown}}, \bibinfo {author} {\bibfnamefont {P.}~\bibnamefont
  {Thompson}}, \bibinfo {author} {\bibfnamefont {J.~A.}\ \bibnamefont {Duffy}},
  \bibinfo {author} {\bibfnamefont {J.~W.}\ \bibnamefont {Taylor}}, \ and\
  \bibinfo {author} {\bibfnamefont {M.~J.}\ \bibnamefont {Cooper}},\ }\href
  {\doibase 10.1107/S0909049502018654} {\bibfield  {journal} {\bibinfo
  {journal} {J. Synchrotron Radiat.}\ }\textbf {\bibinfo {volume} {10}},\
  \bibinfo {pages} {172} (\bibinfo {year} {2003})}\BibitemShut {NoStop}%
\bibitem [{\citenamefont {Klewe}\ \emph {et~al.}(2014)\citenamefont {Klewe},
  \citenamefont {Meinert}, \citenamefont {Boehnke}, \citenamefont {Kuepper},
  \citenamefont {Arenholz}, \citenamefont {Gupta}, \citenamefont {Schmalhorst},
  \citenamefont {Kuschel},\ and\ \citenamefont {Reiss}}]{Klewe:2014}%
  \BibitemOpen
  \bibfield  {author} {\bibinfo {author} {\bibfnamefont {C.}~\bibnamefont
  {Klewe}}, \bibinfo {author} {\bibfnamefont {M.}~\bibnamefont {Meinert}},
  \bibinfo {author} {\bibfnamefont {A.}~\bibnamefont {Boehnke}}, \bibinfo
  {author} {\bibfnamefont {K.}~\bibnamefont {Kuepper}}, \bibinfo {author}
  {\bibfnamefont {E.}~\bibnamefont {Arenholz}}, \bibinfo {author}
  {\bibfnamefont {A.}~\bibnamefont {Gupta}}, \bibinfo {author} {\bibfnamefont
  {J.-M.}\ \bibnamefont {Schmalhorst}}, \bibinfo {author} {\bibfnamefont
  {T.}~\bibnamefont {Kuschel}}, \ and\ \bibinfo {author} {\bibfnamefont
  {G.}~\bibnamefont {Reiss}},\ }\href
  {http://scitation.aip.org/content/aip/journal/jap/115/12/10.1063/1.4869400}
  {\bibfield  {journal} {\bibinfo  {journal} {J. Appl. Phys.}\ }\textbf
  {\bibinfo {volume} {115}},\ \bibinfo {eid} {123903} (\bibinfo {year}
  {2014})}\BibitemShut {NoStop}%
\bibitem [{\citenamefont {Sola}\ \emph {et~al.}(2015)\citenamefont {Sola},
  \citenamefont {Kuepferling}, \citenamefont {Basso}, \citenamefont {Pasquale},
  \citenamefont {Kikkawa}, \citenamefont {Uchida},\ and\ \citenamefont
  {Saitoh}}]{Sola:2015}%
  \BibitemOpen
  \bibfield  {author} {\bibinfo {author} {\bibfnamefont {A.}~\bibnamefont
  {Sola}}, \bibinfo {author} {\bibfnamefont {M.}~\bibnamefont {Kuepferling}},
  \bibinfo {author} {\bibfnamefont {V.}~\bibnamefont {Basso}}, \bibinfo
  {author} {\bibfnamefont {M.}~\bibnamefont {Pasquale}}, \bibinfo {author}
  {\bibfnamefont {T.}~\bibnamefont {Kikkawa}}, \bibinfo {author} {\bibfnamefont
  {K.}~\bibnamefont {Uchida}}, \ and\ \bibinfo {author} {\bibfnamefont
  {E.}~\bibnamefont {Saitoh}},\ }\href@noop {} {\bibfield  {journal} {\bibinfo
  {journal} {J. Appl. Phys.}\ }\textbf {\bibinfo {volume} {117}},\ \bibinfo
  {pages} {17C510} (\bibinfo {year} {2015})}\BibitemShut {NoStop}%
\bibitem [{\citenamefont {Sola}\ \emph {et~al.}(2017)\citenamefont {Sola},
  \citenamefont {Bougiatioti}, \citenamefont {Kuepferling}, \citenamefont
  {Meier}, \citenamefont {Reiss}, \citenamefont {Pasquale}, \citenamefont
  {Kuschel},\ and\ \citenamefont {Basso}}]{Sola:2016}%
  \BibitemOpen
  \bibfield  {author} {\bibinfo {author} {\bibfnamefont {A.}~\bibnamefont
  {Sola}}, \bibinfo {author} {\bibfnamefont {P.}~\bibnamefont {Bougiatioti}},
  \bibinfo {author} {\bibfnamefont {M.}~\bibnamefont {Kuepferling}}, \bibinfo
  {author} {\bibfnamefont {D.}~\bibnamefont {Meier}}, \bibinfo {author}
  {\bibfnamefont {G.}~\bibnamefont {Reiss}}, \bibinfo {author} {\bibfnamefont
  {M.}~\bibnamefont {Pasquale}}, \bibinfo {author} {\bibfnamefont
  {T.}~\bibnamefont {Kuschel}}, \ and\ \bibinfo {author} {\bibfnamefont
  {V.}~\bibnamefont {Basso}},\ }\href@noop {} {\bibfield  {journal} {\bibinfo
  {journal} {arXiv:1701.03285}\ } (\bibinfo {year} {2017})}\BibitemShut
  {NoStop}%
\bibitem [{\citenamefont {Schulz}\ and\ \citenamefont
  {Hoffmann}(2002)}]{Schulz:2002}%
  \BibitemOpen
  \bibfield  {author} {\bibinfo {author} {\bibfnamefont {B.}~\bibnamefont
  {Schulz}}\ and\ \bibinfo {author} {\bibfnamefont {M.}~\bibnamefont
  {Hoffmann}},\ }\href {\doibase 10.1068/htjr018} {\bibfield  {journal}
  {\bibinfo  {journal} {High Temp. - High Pressures}\ }\textbf {\bibinfo
  {volume} {34}},\ \bibinfo {pages} {203} (\bibinfo {year} {2002})}\BibitemShut
  {NoStop}%
\bibitem [{\citenamefont {Nelson}\ \emph {et~al.}(2014)\citenamefont {Nelson},
  \citenamefont {White}, \citenamefont {Andersson}, \citenamefont {Aguiar},
  \citenamefont {McClellan}, \citenamefont {Byler}, \citenamefont {Short},\
  and\ \citenamefont {Stanek}}]{Nelson:2014}%
  \BibitemOpen
  \bibfield  {author} {\bibinfo {author} {\bibfnamefont {A.~T.}\ \bibnamefont
  {Nelson}}, \bibinfo {author} {\bibfnamefont {J.~T.}\ \bibnamefont {White}},
  \bibinfo {author} {\bibfnamefont {D.~A.}\ \bibnamefont {Andersson}}, \bibinfo
  {author} {\bibfnamefont {J.~A.}\ \bibnamefont {Aguiar}}, \bibinfo {author}
  {\bibfnamefont {K.~J.}\ \bibnamefont {McClellan}}, \bibinfo {author}
  {\bibfnamefont {D.~D.}\ \bibnamefont {Byler}}, \bibinfo {author}
  {\bibfnamefont {M.~P.}\ \bibnamefont {Short}}, \ and\ \bibinfo {author}
  {\bibfnamefont {C.~R.}\ \bibnamefont {Stanek}},\ }\href {\doibase
  10.1111/jace.12901} {\bibfield  {journal} {\bibinfo  {journal} {J. Am. Ceram.
  Soc.}\ }\textbf {\bibinfo {volume} {97}},\ \bibinfo {pages} {1559} (\bibinfo
  {year} {2014})}\BibitemShut {NoStop}%
\bibitem [{\citenamefont {Kuschel}\ \emph {et~al.}(2016)\citenamefont
  {Kuschel}, \citenamefont {Klewe}, \citenamefont {Bougiatioti}, \citenamefont
  {Kuschel}, \citenamefont {Wollschläger}, \citenamefont {Bouchenoire},
  \citenamefont {Brown}, \citenamefont {Schmalhorst}, \citenamefont {Meier},\
  and\ \citenamefont {Reiss}}]{Kuschel:2016}%
  \BibitemOpen
  \bibfield  {author} {\bibinfo {author} {\bibfnamefont {T.}~\bibnamefont
  {Kuschel}}, \bibinfo {author} {\bibfnamefont {C.}~\bibnamefont {Klewe}},
  \bibinfo {author} {\bibfnamefont {P.}~\bibnamefont {Bougiatioti}}, \bibinfo
  {author} {\bibfnamefont {O.}~\bibnamefont {Kuschel}}, \bibinfo {author}
  {\bibfnamefont {J.}~\bibnamefont {Wollschläger}}, \bibinfo {author}
  {\bibfnamefont {L.}~\bibnamefont {Bouchenoire}}, \bibinfo {author}
  {\bibfnamefont {S.~D.}\ \bibnamefont {Brown}}, \bibinfo {author}
  {\bibfnamefont {J.~M.}\ \bibnamefont {Schmalhorst}}, \bibinfo {author}
  {\bibfnamefont {D.}~\bibnamefont {Meier}}, \ and\ \bibinfo {author}
  {\bibfnamefont {G.}~\bibnamefont {Reiss}},\ }\href {\doibase
  10.1109/TMAG.2015.2512040} {\bibfield  {journal} {\bibinfo  {journal} {IEEE
  Trans. Magn.}\ }\textbf {\bibinfo {volume} {52}},\ \bibinfo {pages} {4500104}
  (\bibinfo {year} {2016})}\BibitemShut {NoStop}%
\bibitem [{\citenamefont {Meinert}\ and\ \citenamefont
  {Reiss}(2014)}]{Meinert:2014}%
  \BibitemOpen
  \bibfield  {author} {\bibinfo {author} {\bibfnamefont {M.}~\bibnamefont
  {Meinert}}\ and\ \bibinfo {author} {\bibfnamefont {G.}~\bibnamefont
  {Reiss}},\ }\href {http://stacks.iop.org/0953-8984/26/i=11/a=115503}
  {\bibfield  {journal} {\bibinfo  {journal} {J. Phys.: Condens. Matter}\
  }\textbf {\bibinfo {volume} {26}},\ \bibinfo {pages} {115503} (\bibinfo
  {year} {2014})}\BibitemShut {NoStop}%
\bibitem [{\citenamefont {Bougiatioti}\ \emph {et~al.}(2017)\citenamefont
  {Bougiatioti}, \citenamefont {Manos}, \citenamefont {Klewe}, \citenamefont
  {Meier}, \citenamefont {Schmalhorst}, \citenamefont {Kuschel},\ and\
  \citenamefont {Reiss}}]{Bougiatioti:2017}%
  \BibitemOpen
  \bibfield  {author} {\bibinfo {author} {\bibfnamefont {P.}~\bibnamefont
  {Bougiatioti}}, \bibinfo {author} {\bibfnamefont {O.}~\bibnamefont {Manos}},
  \bibinfo {author} {\bibfnamefont {C.}~\bibnamefont {Klewe}}, \bibinfo
  {author} {\bibfnamefont {D.}~\bibnamefont {Meier}}, \bibinfo {author}
  {\bibfnamefont {J.-M.}\ \bibnamefont {Schmalhorst}}, \bibinfo {author}
  {\bibfnamefont {T.}~\bibnamefont {Kuschel}}, \ and\ \bibinfo {author}
  {\bibfnamefont {G.}~\bibnamefont {Reiss}},\ }\href@noop {} {\bibfield
  {journal} {\bibinfo  {journal} {in preparation}\ } (\bibinfo {year}
  {2017})}\BibitemShut {NoStop}%
\bibitem [{\citenamefont {Lord}\ and\ \citenamefont
  {Parker}(1960)}]{Lord:1960}%
  \BibitemOpen
  \bibfield  {author} {\bibinfo {author} {\bibfnamefont {H.}~\bibnamefont
  {Lord}}\ and\ \bibinfo {author} {\bibfnamefont {R.}~\bibnamefont {Parker}},\
  }\href@noop {} {\bibfield  {journal} {\bibinfo  {journal} {Nature}\ }\textbf
  {\bibinfo {volume} {188}},\ \bibinfo {pages} {929} (\bibinfo {year}
  {1960})}\BibitemShut {NoStop}%
\bibitem [{\citenamefont {Klewe}\ \emph {et~al.}(2016)\citenamefont {Klewe},
  \citenamefont {Kuschel}, \citenamefont {Schmalhorst}, \citenamefont
  {Bertram}, \citenamefont {Kuschel}, \citenamefont {Wollschl\"ager},
  \citenamefont {Strempfer}, \citenamefont {Meinert},\ and\ \citenamefont
  {Reiss}}]{Klewe:2016}%
  \BibitemOpen
  \bibfield  {author} {\bibinfo {author} {\bibfnamefont {C.}~\bibnamefont
  {Klewe}}, \bibinfo {author} {\bibfnamefont {T.}~\bibnamefont {Kuschel}},
  \bibinfo {author} {\bibfnamefont {J.-M.}\ \bibnamefont {Schmalhorst}},
  \bibinfo {author} {\bibfnamefont {F.}~\bibnamefont {Bertram}}, \bibinfo
  {author} {\bibfnamefont {O.}~\bibnamefont {Kuschel}}, \bibinfo {author}
  {\bibfnamefont {J.}~\bibnamefont {Wollschl\"ager}}, \bibinfo {author}
  {\bibfnamefont {J.}~\bibnamefont {Strempfer}}, \bibinfo {author}
  {\bibfnamefont {M.}~\bibnamefont {Meinert}}, \ and\ \bibinfo {author}
  {\bibfnamefont {G.}~\bibnamefont {Reiss}},\ }\href {\doibase
  10.1103/PhysRevB.93.214440} {\bibfield  {journal} {\bibinfo  {journal} {Phys.
  Rev. B}\ }\textbf {\bibinfo {volume} {93}},\ \bibinfo {pages} {214440}
  (\bibinfo {year} {2016})}\BibitemShut {NoStop}%
\end{thebibliography}%

\end{document}